\documentstyle[psfig]{l-aa}

\begin{document}

\thesaurus{10.11.1,10.19.2}

\title{3D self-consistent $N$-body barred models of the Milky Way}
\subtitle{I. Stellar dynamics}

\author{R. Fux}

\institute{Geneva Observatory, Ch. des Maillettes 51, CH-1290 Sauverny,
Switzerland}

\offprints{R. Fux}

\date{Received April 23; accepted June 20, 1997}
\maketitle

\begin{abstract}
Many 3D $N$-body barred models of the Galaxy extending beyond the Solar circle
are realised by self-consistent evolution of various bar unstable axisymmetric
models. The COBE/DIRBE $K$-band map, corrected for extinction, is used to
constrain the location of the observer in these models, assuming a constant
mass-to-light ratio. The resulting view points in the best matching models
suggest that the inclination angle of the Galactic bar relative to the
Sun-Galactic centre line is $28\degr\pm 7\degr$.
\par Scaling the masses according to the observed radial velocity dispersion
of M giants in Baade's Window, several models reproduce satisfactorily the
kinematics of disc and halo stars in the Solar neighbourhood, as well as the
disc local surface density and scale parameters. These models have a face-on
bar axis ratio $b/a=0.5\pm 0.1$ and a bar pattern speed $\Omega_{\rm P}=
50\pm 5$~km/s/kpc, corresponding to a corotation radius of $4.3\pm 0.5$~kpc.
The HI terminal velocity constraints favour models with low disc mass fraction
near the centre.
\par The large microlensing optical depths observed towards the Galactic bulge
exclude models with a disc scale height $h_z\la 250$~pc around $R=4$~kpc,
arguing for a constant thickness Galactic disc. The models also indicate that
a spiral arm starting at the near end of the bar can contribute as much as
$0.5\times 10^{-6}$ to the optical depth in Baade's Window.
The mass-to-$K$ luminosity ratio of the Galactic bulge is probably more than
0.7 (Solar units), and if the same ratio applies outside the bar region,
then the Milky Way should have a maximum disc.

\vspace{-.6cm}
\end{abstract}
\vspace{-.4cm}
%
\section{Introduction}
Recent results demonstrate that the Milky Way, as more than 2/3 of disc
galaxies and as suspected early on by de~Vaucouleurs (\cite{DV}) from a
comparison of gas kinematics towards the Galactic centre and in external
galaxies, is a barred galaxy with the near side of the bar pointing in the
first Galactic quadrant. Evidence comes from near-IR surface photometry,
discrete source counts, gas and stellar kinematics and gravitational
microlensing (see Kuijken \cite{KU2} and Gerhard \cite{G} for reviews). The
most suggestive data certainly are the COBE/DIRBE near-IR maps of the Galactic
bulge (Weiland et al. \cite{WA}; Dwek et al. \cite{DA}; Binney et al.
\cite{BGS} for a non-parametric deprojection). Estimates of the angle between
the major axis of the bar and the Sun-Galactic centre line range from
$10\degr$~to~$45\degr$ (e.g. Stanek et al. \cite{SU}).
\par Furthermore, the distribution of HII regions, young stellar clusters,
HI gas, CO clouds and dust betray the existence of several Galactic spiral
arms (Mihalas \& Binney \cite{MB}; Vall\'ee \cite{V} and reference therein),
and external spiral galaxies of the same (Sbc) Hubble type as the Milky Way
have arm-interarm surface mass density ratios of order 2 out to at least 3
disc scale lengths from the centre (Rix \& Zaritsky \cite{RZ}). Hence the
detailed structure of our Galaxy clearly deviates from axisymmetry.
\par No self-consistent 3D dynamical barred model of the Galaxy including
simultaneously stellar, gas and dark components and extending beyond the Sun's
Galactocentric distance has been proposed yet. Existing stellar dynamical
models are either axisymmetric (Kuijken \& Dubinski \cite{KD}; Durand et al.
\cite{DD}) and/or restricted to a single Galactic component (Zhao \cite{Z} for
the COBE-bar; Kent \cite{K} and Kuijken \cite{KU1} for an axisymmetric bulge
model), and gas flow calculations always assume a rigid rotating bar potential
(Mulder \& Liem \cite{ML}; Wada \cite{WT}; Weiner \& Sellwood \cite{WS}).
\par This paper presents the first step of a program aiming such a complete
model. Many self-consistent pure stellar dynamical barred models of the Milky
Way are built from $N$-body evolution of bar unstable axisymmetric models.
This method, already applied to the Galaxy by Sellwood (\cite{S1}; \cite{S2}),
naturally takes into account the main dynamical processes acting in the
evolution of real isolated galaxies, like those responsible for spontaneous
bar formation and self-sustained spiral structures (e.g. Zhang \cite{ZX}). The
$N$-body method also proves convenient to cope with a dissipative gas
component, as will be added to the models in the next step and described in
a second paper.
\par The structure of this paper is as follow: in~Sect.~2 we describe the
distinct initial conditions of the various components considered in the
simulations. In Sect.~3, we give some technical informations about the
$N$-body integration and present the time evolution of the initial models.
In Sect.~4 we determine for each evolved model the best location of the
observer (the Sun) according to the COBE/DIRBE near-IR observations of the
Galactic bulge. In Sect.~5 we fix the velocity scales of the models to match
the observed stellar velocity dispersion in Baade's Window and discuss some
of the resulting absolute model properties.
%
\section{Initial conditions}
Equilibrium or close to equilibrium axisymmetric phase space density functions
(DF) have been obtained for disc galaxies either by applying the strong Jeans
theorem which states that the DF is an explicit function of at most three
isolating independent integrals of motion (Kuijken \& Dubinski \cite{KD};
Durand et al. \cite{DD}), or by solving the hydrodynamical Jeans equations for
the velocity moments under some arbitrary closure conditions and assuming a
Gaussian velocity distribution (Hernquist \cite{HL}).
\par The first method has the advantage to provide exact solutions of the
Boltzmann equation, but is in general (except for St\"ackel potentials)
limited by the lack of an analytical third integral: DFs depending only on the
two classical integrals, i.e. the total energy and the angular momentum about
the symmetry axis ($z$), always have $\sigma_{Rz}^2=0$, unsuitable for
substantially anisotropic spheroidal components, and $\sigma_{zz}^2=
\sigma_{RR}^2$, which in the Solar neighbourhood is wrong for any evolved
stellar population. Moreover, generating DFs with imposed mass densities
requires the cumbersome inversion of the integral equation that relates these
two quantities (Kuijken \cite{KU1} and references therein).
\par The second method allows for a larger variety of velocity ellipsoids,
depending on the closure conditions, and is also more adapted for specified
mass distributions. It was therefore retained here, with some modification
regarding the shape of the velocity distribution.
\par The initial models are described in standard astronomical units, assuming
that the Galactocentric distance of the Sun is $R_{\circ}=8$~kpc. These units
will serve as reference in the evolved models until~Sect.~\ref{disc}.
\subsection{Mass distribution}
The initial mass density $\rho$ in our simulations includes three axisymmetric
components. 
\par The first one is an oblate stellar nucleus-spheroid (NS) inspired from
the model of Sellwood \& Sanders (\cite{SS}):
\begin{equation}
\rho_{\rm NS}(s) = \frac{M_{\rm NS}}{4\pi a^3 e I_\infty}
                   \cdot \frac{(s/a)^{p}}{1+(s/a)^{p-q}},
\label{rhoNS}
\end{equation}
where
\begin{eqnarray}
\hspace{1cm} s^2 & \equiv & R^2+z^2/e^2, \label{s} \\
I_\infty & = &\frac{\pi}{p-q} \csc\left[\frac{(p+3)\pi}{p-q}\right],
\label{m2}
\end{eqnarray}
$a$ is a knee radius, $e$ the axis ratio and $M_{\rm NS}$ the integrated mass.
Setting $p=-1.8$ and $q=-3.3$, this mass density behaves as $s^{-1.8}$ for
$s\ll~a$, in agreement with near-IR observations of the Galactic inner kpc
(Becklin \& Neugebauer \cite{BN}; Matsumoto et al. \cite{MH}) if a constant
mass-to-light ratio is assumed, and as $s^{-3.3}$ for $s\gg~a$, similar to the
radial number density decrease of RR Lyrae stars (Preston et al. \cite{PS})
and of the globular clusters (Zinn \cite{ZI}). This component is therefore
well suited to represent the nuclear bulge and the stellar halo.
\par The second component is a double exponential stellar disc:
\begin{equation}
\rho_{\rm D}(R,z) = \frac{M_{\rm D}}{4\pi h_R^2 h_z}
                    \cdot \exp\left[-\frac{R}{h_R}-\frac{|z|}{h_z}\right],
\label{rhoD}
\end{equation}
where $h_R$, $h_z$ and $M_{\rm D}$ are resp. the scale length, the
scale height and the integrated mass. This component stands here for the
Galactic old disc.
\par Finally an oblate exponential dark halo (DH) with the same axis ratio $e$
as the NS component is added to ensure a flat rotation curve at large radii:
\begin{equation}
\rho_{\rm DH}(s) = \frac{M_{\rm DH}}{8\pi b^3 e}\exp(-s/b),
\label{rhoDH}
\end{equation}
where $b$ is the scale length and $M_{\rm DH}$ the total dark mass.
\subsection{Truncation and flattening}
To minimise the number of particles outside the force grid and increase the
particle statistics near the centre (at fixed number of particles), we softly
truncate each component multiplying its mass density by:
\begin{equation}
f(s) = \tanh\left[\frac{s-R_{\rm c}}{2 \delta}\right],
\label{trunc}
\end{equation}
where $R_{\rm c}$ is the truncation radius and $\delta$ the width over which
$f$ falls from 1 to 0. The densities thus vanish on the spheroidal surface
$s=R_{\rm c}$.
\par In all simulations we set $R_{\rm c}=38$~kpc, $\delta=5$~kpc and $e=0.5$.
The choice of $e$ is realistic at least for the NS component and has the
technical advantage over more spherical models to limit the $z$-extension of
the force grid.
\subsection{Choice of mass density parameters}
The first row in Table~\ref{init} lists the values of the parameters $a$,
$M_{\rm NS}$, $h_R$, $h_z$, $M_{\rm D}$, $b$ and $M_{\rm DH}$ that were
adopted for the ``central'' simulation~m00. The choice of $a$ and $M_{\rm NS}$
resp. refer to the transition radius where the Galactic near-IR
emissivity starts to deviate significantly from the inner power law, and to a
mass-to-light ratio in the nucleus region of 0.75 Solar units in the $K$-band,
corresponding to a compromise between the values of 0.5 derived by Binney et
al. (\cite{BG}) from gas dynamics and 1 in Kent's (\cite{K}) dynamical bulge
model. The disc scales $h_R$ and $h_z$ are based on recent determinations
(Fux \& Martinet \cite{FM}; Ruphy et al. \cite{RR}; Sackett \cite{S} for a
review) and are also consistent with near-IR photometry (Kent et al.
\cite{KDF}).
\par
\begin{figure}[t]
\centerline{\psfig{figure=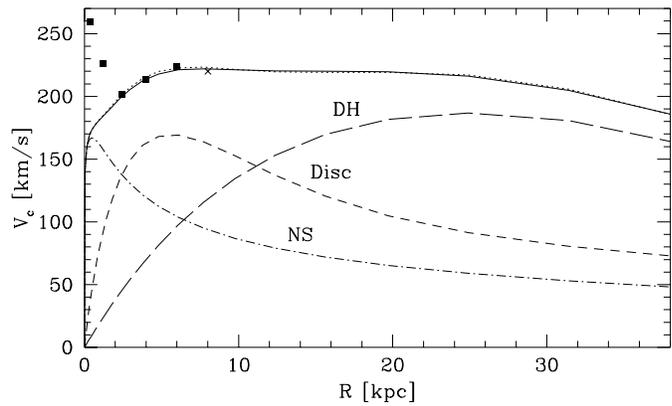,width=8.8cm}}
\caption[]{Rotation curve of the initial m00~model, with the contributions of
         each component. The squares are tangent point circular velocities
         based on the HI terminal velocities compiled by Caldwell \& Ostriker
         (\cite{CO}), assuming $R_{\circ}=8$~kpc and $V_{\circ}=220$~km/s, as
         indicated by the cross. The almost coinciding full and dotted lines
         represent resp. the circular velocities before and after
         relaxing the DH component}
\label{crot}
\end{figure}

 The last three parameters $M_{\rm D}$, $b$ and $M_{\rm DH}$ were adjusted
to reproduce the Galactic rotation curve beyond 3~kpc, as illustrated in
Fig.~\ref{crot}, while holding the local surface mass density of the disc into
a reasonable range. The inner peak of the observed rotation curve is an 
artifact due to non-circular gas motion in the bar region and therefore does
not need to be reproduced in axisymmetric models. The outer rotation curve of
the resulting initial m00~model is fairly flat inside 25~kpc and then starts
to decrease because of the exponential decline of the DH density and the mass
truncation.
\par At $R=R_{\circ}$, the disc and total disc+NS+DH surface densities are
resp. 48~$M_{\odot}$/pc$^2$ (integrated over all $z$) and
66~$M_{\odot}$/pc$^2$ within $|z| < 700$~pc, and the disc, NS and DH volume
densities resp. 0.095, $1.3\cdot 10^{-3}$ and 0.014~$M_{\odot}$/pc$^3$,
summing up to give $\rho_{\rm tot}(R_{\circ},0)=0.11$~$M_{\odot}$/pc$^3$.
Our local spheroid volume density exceeds estimates from low metallicity and
high velocity stars (Bahcall et al. \cite{BSS}). Part of the excess could be
attributed to a missing thick disc component in the model.
\par The initial conditions need not to mimic precisely the Milky Way because
the bar instability expected during time evolution will drastically affect the
phase-space distribution. Since we do not control a priori the issue of a
simulation, many runs are required to appreciate the effect of the initial
parameters on the final properties of the evolved models. For this reason
we have realised, in addition to the simulation~m00, ten further simulations
where each of the parameters $a$, $M_{\rm NS}$, $h_R$, $h_z$ and $M_{\rm D}$
has been separately set to a lower and higher value than in m00. The adopted
values and nomenclature for each simulation are listed in Table~\ref{init}.
The DH parameters were always kept at the same values.
\begin{table}
\centering
\caption[]{Mass density parameters of the initial models. Distances are in~kpc
         and masses in $10^{10}$~$M_{\odot}$. Model~m00 is about centred in
         the explored parameter space. Unspecified values in the other models
         are similar to those in m00. Model~m11 differ from m00 only by the DH
         kinematics and model~m12 by the subsequent symmetry-free time
         integration. The parameters $M_{\rm NS}$ and $M_{\rm DH}$ are masses
         without the truncation by Eq.~(\ref{trunc})}
\begin{tabular}{cccccccc} \hline
Model & $a$ & $M_{\rm NS}$ & $h_R$ & $h_z$ & $M_{\rm D}$ & $b$ & $M_{\rm DH}$
                                                                    \\ \hline
m00 & 1.0 & 3.0 & 2.5 & 0.25 & 4.6 & 9.1 & 32.0 \\
m01 & 2.0 &  .  &  .  &  .   &  .  &  .  &   .  \\
m02 & 0.5 &  .  &  .  &  .   &  .  &  .  &   .  \\
m03 &  .  & 1.6 &  .  &  .   &  .  &  .  &   .  \\
m04 &  .  & 5.0 &  .  &  .   &  .  &  .  &   .  \\
m05 &  .  &  .  & 2.0 &  .   &  .  &  .  &   .  \\
m06 &  .  &  .  & 4.0 &  .   &  .  &  .  &   .  \\
m07 &  .  &  .  &  .  & 0.13 &  .  &  .  &   .  \\
m08 &  .  &  .  &  .  & 0.35 &  .  &  .  &   .  \\
m09 &  .  &  .  &  .  &  .   & 4.2 &  .  &   .  \\
m10 &  .  &  .  &  .  &  .   & 5.0 &  .  &   .  \\ \hline
m11 & . & . & \multicolumn{3}{c}{non-rotating DH} & . & . \\
m12 & . & . & \multicolumn{3}{c}{no symmetries}   & . & . \\ \hline
\label{init}
\end{tabular}
\end{table}
\subsection{Isotropic Gaussian velocity distribution}
Preliminary simulations (Fux et al. \cite{FD}) with isotropic and Gaussian
initial velocity distribution for each component were performed to check how
self-consistent evolution would rearrange the phase space distribution of such
simple initial conditions.
\par Simultaneous evolution of all mass components first led to unacceptable
strong expanding rings of overdensity in the disc, mainly excited by radial
mass oscillations of the DH damping out over a dynamical time scale. The DH
was therefore individually relaxed during 3~Gyr with imposed axisymmetry before
releasing the other components, suppressing indeed most of the subsequent
perturbations in the disc. The origin of the DH disequilibrium will be
discussed in~Sect.~\ref{DH}.
\par In pre-relaxed DH simulations, the velocity dispersion of the disc
becomes spontaneously anisotropic within a few rotation periods, with a planar
anisotropy compatible with first order epicycle theory (Binney \& Tremaine
\cite{BT}) and ratios between the velocity dispersion components around
$R=R_{\circ}$ similar to those observed in the Solar neighbourhood for the old
disc (see Fig.~2 in Fux et al. \cite{FD} and Table~\ref{loc}). Hence isotropic
initial kinematics seems to suffice for the disc.
\par However, the kinematics of the NS never reaches the observed radial
velocity dispersion anisotropy of the local Galactic halo stars and instead
sustains gravity through too fast rotation, exceeding half the circular
velocity (see Fig.~\ref{cinNS}c and Table~\ref{loc}). The mean rotation
velocity can of course be arbitrarily reduced by changing the sign of the
azimuthal velocities of selected particles, but then the rotation velocity
dispersion would increase and in turn deviate from the local observations.
Some radial anisotropy should therefore appear already in the initial velocity
distribution of the NS component.
\subsection{Anisotropic velocity dispersion}
\label{ani}
To achieve higher degree of radial velocity dispersion anisotropy in both the
NS and the DH components, we have solved the Jeans equations for more general
closure conditions than just isotropy.
\par Following Bacon et al. (\cite{BSM}), we assume that the velocity
ellipsoid points everywhere towards the Galactic centre,
i.e. $\sigma_{r\theta}^2=0$ in spherical coordinates, with a free anisotropy
parameter $\beta \equiv 1-\sigma_{\theta \theta}^2/\sigma_{rr}^2$ depending
only on $r$ and of the form:
\begin{equation}
\beta(r) = \beta_{\infty} \frac{r}{\sqrt{r^2+r_{\circ}^2}},
\label{betar}
\end{equation}
where $r_{\circ}$ is a transition radius and $\beta_{\infty}$ the asymptotic
anisotropy at large $r$: $\beta \propto r$ for $r \ll r_{\circ}$, and $\beta=
\beta_{\infty}$ for $r \gg r_{\circ}$. We also assume no other streaming motion
than rotation about the symmetry axis, i.e.
$\overline{v_r}=\overline{v_{\theta}}=0$, and take as boundary conditions
$\sigma_{rr}^2=\sigma_{\theta \theta}^2=0$ on the mass truncation surface.
The details of the numerical method, which can in fact also handle anisotropy
parameter depending on $\theta$, are presented in Fux (\cite{F}).
\par The solutions provide
$\overline{v_{\phi}^2}=\overline{v_{\phi}}^2+\sigma_{\phi \phi}^2$, leaving
free the relative contributions of organised and random velocity in the
$\phi$ direction. As a convenient choice, which ensures isotropy near the 
centre and low rotation for $r \gg r_{\circ}$, we~set:
\begin{equation}
\sigma_{\phi \phi}^2 = \frac{r_{\circ}^2 \sigma_{\theta \theta}^2
                      +r^2 \overline{v_\phi^2}}{r_{\circ}^2+r^2},
\label{spp}
\end{equation}
where $r_{\circ}$ is the same parameter as in Eq.~(\ref{betar}).
\par
\begin{figure}[t]
\centerline{\psfig{figure=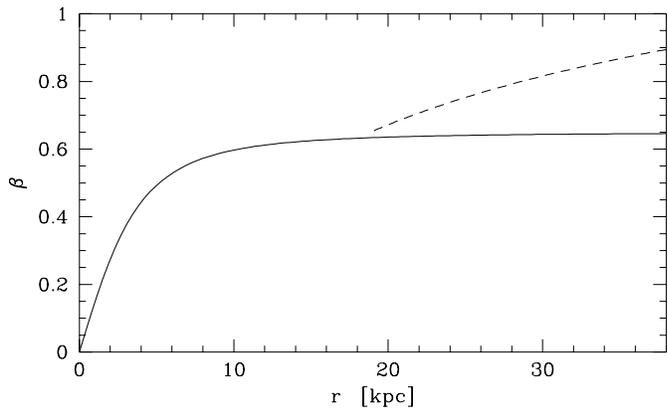,width=8.8cm}}
\caption[]{Full line: anisotropy parameter $\beta(r)$ of the NS component in
all initial models. Dashed line: upper limit for positive
$\overline{v_{\phi}^2}$ on the mass truncation surface (in model~m00)}
\label{bet}
\end{figure}

 The values of $r_{\circ}$ and $\beta_{\infty}$ are restricted by the
condition $\overline{v_{\phi}^2}\geq 0$ everywhere. In particular, on the
truncation surface of our initial mass models, this condition imposes an upper
limit to $\beta(r)$ in the range $eR_{\rm c}<r<R_{\rm c}$ depending only on
the potential and its first derivatives (see Fig.~\ref{bet}). The resulting 
strongest constraint is $\beta(eR_{\rm c})<0.65$.
\begin{table}[t]
\centering
\caption[]{Observed properties of the stellar halo (subdwarfs) and the old
disc in the Solar neighbourhood. $\Sigma_{\circ}$ is the $total$ thin disc
surface density (i.e. including also the young disc). The values of the disc
velocity dispersion are averages over $z$. References are: (1)~Majewski
\cite{M}, (2)~Wielen \cite{WR}, (3)~Sackett \cite{S}, (4)~Kuijken \& Gilmore
\cite{KG}}
\begin{tabular}{llcc} \hline
  &&& Reference \\
  & \vspace{-.3cm} & \\
Subdwarfs:    & $\sigma_{rr}=\sigma_{RR}$            & $131\pm  6$ km/s & 1 \\
              & $\sigma_{\phi \phi}$                 & $106\pm  6$ km/s & 1 \\
              & $\sigma_{\theta \theta}=\sigma_{zz}$ &  $85\pm  4$ km/s & 1 \\
              & $\overline{v_{\phi}}$                &  $37\pm 10$ km/s & 1 \\
  & \vspace{-.3cm} & \\
Old disc:     & $\sigma_{RR}$                        &  $48\pm  3$ km/s & 2 \\
              & $\sigma_{\phi \phi}$                 &  $29\pm  2$ km/s & 2 \\
              & $\sigma_{zz}$                        &  $25\pm  2$ km/s & 2 \\
              & $h_z$                                & $300\pm 25$ pc   & 3 \\
  & \vspace{-.3cm} & \\
Thin disc:    & $\Sigma_{\circ}$                     &  $48\pm  8$ 
                                               $M_{\odot}$/pc$^2$ & 4 \\ \hline
\end{tabular}
\label{loc}
\end{table}
\subsection{Dark halo disequilibrium and velocity distribution}
\label{DH}
Coming back to the DH radial oscillations, single relaxation of DHs
with Gaussian initial velocity distributions but variable radial velocity
dispersion anisotropy based on the technique outlined above (Sect.~\ref{ani})
indicate that the oscillations strengthen with the amount of anisotropy.
From these experiments we inferred that the mass oscillations are in fact a
consequence of the Gaussian tails in the velocity distribution and the finite
DH mass extent: a significant fraction of the DH particles, about 10\% in the
isotropic case and increasing with radial anisotropy, have velocities which
carry them outside the truncation surface, unbalancing therefore the inner
equilibrium.
\par To overcome this problem, the Gaussian velocity distributions of the
extended DH {\it and} NS components have been replaced by a bounded 3D
distribution build upon standard Beta distributions. This new
\mbox{``$B_3$''-distribution} has four parameters, $\kappa$, $\lambda$, $\mu$
and $\omega$ (one per adjustable velocity moment), and is described as a
function of its reduced variables $\xi$, $\eta$ and $\zeta$ in Appendix~A.
\par With the substitutions $\xi \equiv v_{\phi}/v_{\circ}$,
$\eta \equiv v_r/v_{\rm e}$ and $\zeta \equiv v_{\theta}/v_{\rm e}$, the
distribution is bounded in velocity space by a spheroidal surface with
principal axes aligned with the $v_{\phi}$, $v_r$ and $v_{\theta}$ axes and of
half-length $v_{\circ}$ along $v_{\phi}$ and $v_{\rm e}$ along $v_r$ and
$v_{\theta}$. The boundary surface is not taken as a sphere because particles
launched at a given spatial position can afford higher velocities in the
tangential direction than in the radial direction without escaping the system.
Indeed, the boundary velocities $v_{\circ}$ and $v_{\rm e}$ can be quantified
using the integrals of motion. If the velocity of a particle is decomposed in
an azimuthal component, $v_{\phi}$, and a meridional component, $v_{\rm m}$,
then its energy writes $E=\frac{1}{2}(v_{\rm m}^2+L_z^2/R^2)+\Phi$, where
$L_z=R\cdot v_{\phi}$ is the angular momentum about the $z$-axis and
$\Phi(R,z)$ is the gravitational potential. Assuming that the mass density is
bounded by an equipotential surface $\Phi_{\rm p}$ and imposing $v_{\rm m}=0$
when a particle reaches this surface (otherwise the particle would cross the
surface and escape), the conservation of energy and angular momentum yield the
following condition for the confinement of the particle orbit inside the 
system:
\begin{equation}
v_{\rm m}^2+\frac{R_{\rm p}^2-R^2}{R_{\rm p}^2}v_{\phi}^2
                    <2\left[\Phi_{\rm p}-\Phi(R,z)\right],
\label{vlim}
\end{equation}
where $R_{\rm p}$ is the maximum cylindrical radius attained by the particle.
Thus:
\begin{eqnarray}
v_{\rm e} & = & \sqrt{2(\Phi_{\rm p}-\Phi)}, \label{ve} \\
v_{\circ} & = & \frac{R_{\rm p}}{\sqrt{R_{\rm p}^2-R^2}} \cdot v_{\rm e},
                                                              \label{vo}
\end{eqnarray}
and $v_{\circ}>v_{\rm e}$.
\par In our initial models, since the truncation surface
of the mass distribution is not an equipotential, we simply take the maximum
value of the potential on this surface, i.e. $\Phi_{\rm p} \equiv
\Phi(R_{\rm c},0)$, and set $R_{\rm p}=R_{\rm c}$. For the initial m00~model,
the resulting behaviour of $v_{\rm e}$ and $v_{\circ}$ in the plane is shown
in Fig.~\ref{cinNS}a. This will not prevent some particles to escape but at
least considerably reduces the DH disequilibrium and ensures well defined
values of the boundary velocities everywhere inside the truncation surface.
\par If $\kappa, \lambda\leq 2$, $\mu\leq 3/2$ or $\omega \leq 1$, the
velocity distribution presents unphysical singularities on the boundary
spheroid which should in principle be avoided. In particular, assuming
$\overline{v_{\phi}}>0$ and substituting the requirement $\lambda>2$ in
Eq.~(\ref{lambda}) yields:
\begin{equation}
\overline{v_{\phi}}>\frac{-(v_{\circ}^2-\overline{v_{\phi}^2})
                    +\sqrt{(v_{\circ}^2-\overline{v_{\phi}^2})
            -16v_{\circ}^2(v_{\circ}^2-5\overline{v_{\phi}^2})}}{8v_{\circ}},
\label{vmin}
\end{equation}
which puts a lower limit on the mean rotation velocity, and thus further
restricts the range of the parameter $r_{\circ}$. For the adopted mass
truncation radius $R_{\rm c}$ and flattening~$e$, this condition unfortunately
prevents NS models with mean rotation velocity below 100~km/s when approaching
$R=R_{\rm c}$ (see Fig.~\ref{cinNS}a). Therefore, to reproduce the observed
rotation of the stellar halo in the Solar neighbourhood with Eq.~(\ref{spp}),
we decided to violate this condition for the NS component and accept irregular
velocity distributions in a minor portion of the $R\!-\!z$ plane.
Similarly, the requirements $\mu>3/2$ and $\omega>1$ put constraints on the
meridional components of the velocity dispersion which could be satisfied in
all NS models, but not in the hot central part of the DHs.
\par The DH pre-relaxations have been maintained even with the adopted $B_3$
velocity distribution to minimise any persistent DH induced
perturbations on the visible components. For these components, the Jeans
equations were solved using the total potential {\it after} relaxation, the
noise in the potential being reduced by averaging the DH mass density in time.
The DH relaxation does not modify much the starting potential
(see Fig.~\ref{crot}).
\subsection{Choice of velocity parameters}
\begin{figure}[t]
\centerline{\psfig{figure=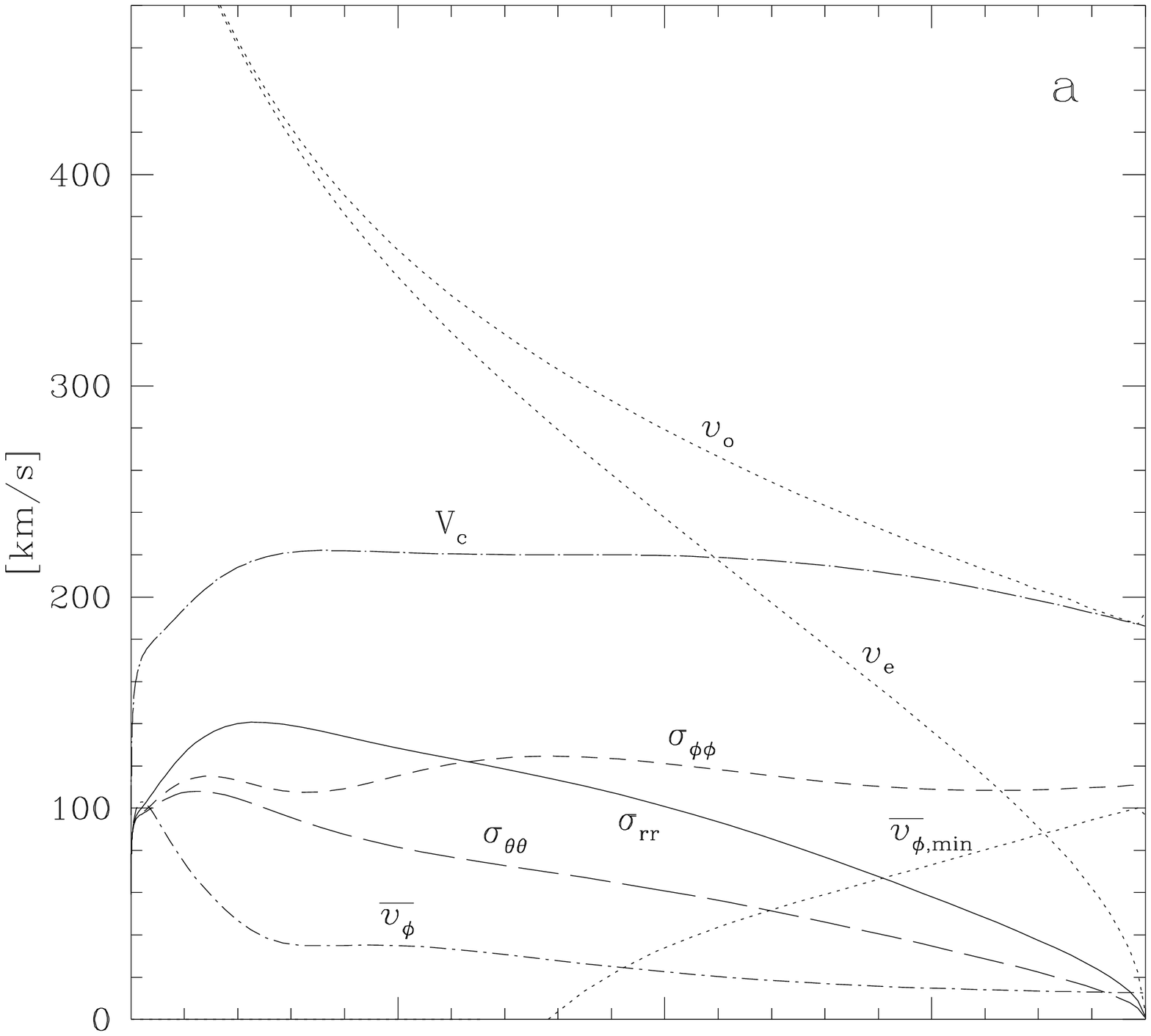,width=8.8cm}}
\centerline{\psfig{figure=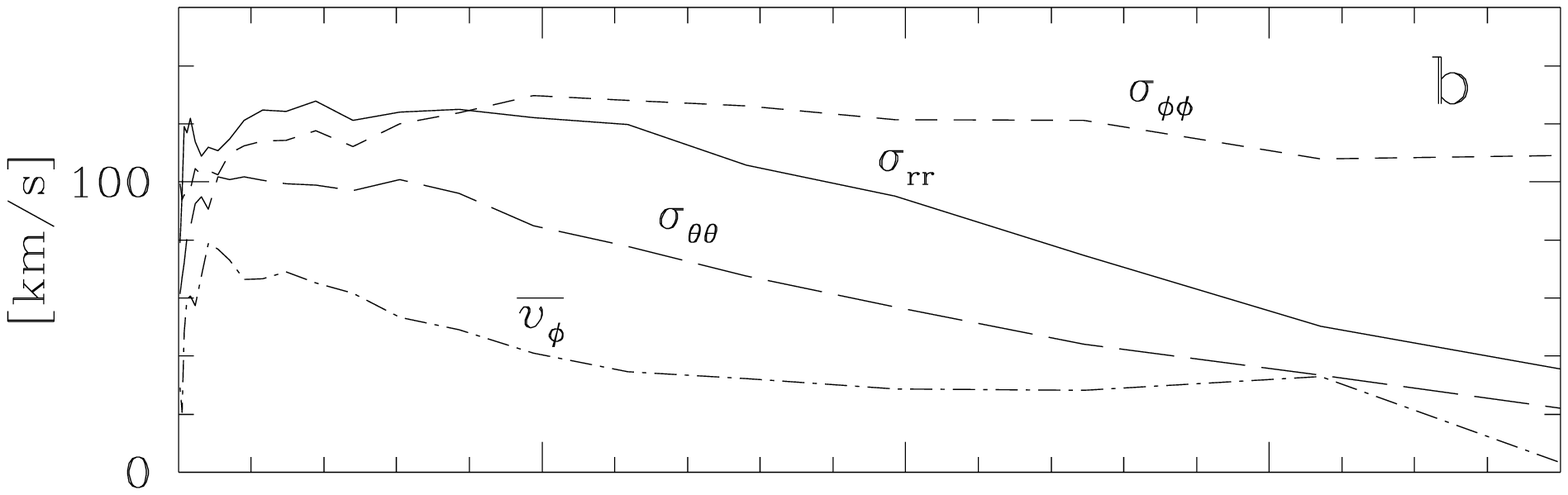,width=8.8cm}}
\centerline{\psfig{figure=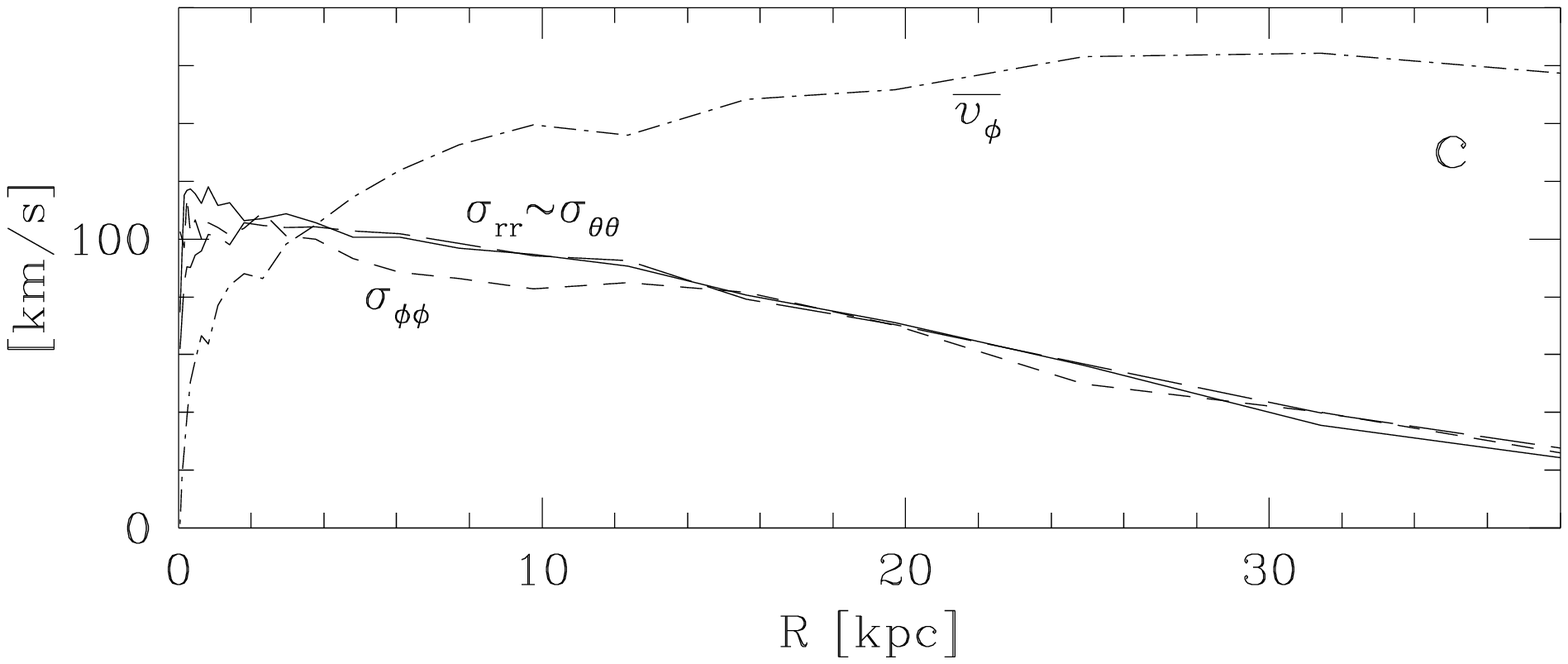,width=8.8cm}}
\caption[]{{\bf a} Initial velocity dispersion and mean rotation velocity of
the NS component at $z=0$ in simulation~m00. The two upper dotted lines give
the meridional $(v_{\rm e})$ and azimuthal $(v_{\circ})$ escape velocities
resulting from Eqs.~(\ref{ve}) and (\ref{vo}), and the lower dotted line the
minimum admissible mean rotation ($\overline{v_{\phi}}$$_{,{\rm min}}$) 
defined in Eq.~(\ref{vmin}) for regular velocity distribution. The circular
velocity $(V_{\rm c})$ is also represented. {\bf b} Corresponding velocity
moments at $t=1200$~Myr. {\bf c} Velocity moments at the same time in a
simulation identical to m00 except that the NS component has isotropic
initial velocity dispersion}
\label{cinNS}
\end{figure}

For the NS of simulation~m00, the parameters $r_{\circ}$ and $\beta_{\infty}$
were adjusted to reproduce the observed rotation and velocity dispersion of
the stellar halo in the Solar neighbourhood given in Table~\ref{loc}.
The adopted parameters are $r_{\circ}=4.3$~kpc and $\beta_{\infty}=0.65$,
leading to the very close to critical $\beta(r)$ profile displayed in
Fig.~\ref{bet}. The resulting initial velocity moments of the NS are shown in
Fig.~\ref{cinNS}a.
\par The reversal of the $\sigma_{rr}^2$ versus $\sigma_{\phi \phi}^2$
anisotropy at $R\approx 13$~kpc is consistent with the kinematics of the blue
horizontal branch field stars (Sommer-Larsen et al. \cite{SF}) and other halo
stars (Beers \& Sommer-Larsen \cite{BSL}). The azimuthal anisotropy at large
radii could reflect the fact that stars easier escape in the radial direction,
and that $\sigma_{rr}^2$ must vanish on the boundary of a finite and
stationary system, whereas $\sigma_{\phi \phi}^2$ can still be supported by
low eccentricity orbits.
\par Figure~\ref{cinNS}b shows the NS kinematics in simulation~m00 shortly
before the formation of the bar. Within $R\la\!14$ kpc, the NS looses a
part of its radial velocity dispersion anisotropy and its rotation velocity
raises. These changes certainly reflect the extreme nature of our initial
conditions, i.e. forcing the radial anisotropy to its maximum. Nevertheless,
the re-adjusted velocity moments still remain much closer to the observations
than those of simulations started with isotropic NS velocity dispersion, as
illustrated in Fig.~\ref{cinNS}c.
\par For the NS of the other simulations, we simply use the same $r_{\circ}$
and $\beta_{\infty}$ values than in simulation~m00. For the DHs of simulations
m00-m10, we set $r_{\circ}=b$ and $\beta_{\infty}=0.1$, compatible with the
restriction of Eq.~(\ref{vmin}). The initial conditions for simulation~m11 are
identical with those of m00, except that the DH has the same anisotropy
$\beta(r)$ as the NS and no net rotation, i.e. $\sigma_{\phi\phi}^2=
\overline{v_{\phi}^2}$ instead of Eq.~(\ref{spp}), and hence also present an
irregular outer \mbox{$v_{\phi}$-distribution}. As justified by the preliminary
simulations, all discs have isotropic ($\beta=0$) and Gaussian initial
velocity distribution, implying a velocity dispersion tightly related to
their scale heights.
%
\section{Time evolution}
The simulations m00-m11 were all done imposing 2-fold rotational symmetry
about the $z$-axis and reflection symmetry about the plane $z=0$, hence
reducing the numerical noise of the potential. For comparison, the
\mbox{initial} m00~model was also integrated without any symmetry, providing
our last m12~simulation. Each simulation is runned up to $t=5$~Gyr, and ouputs
of the particle phase-space coordinates were realised every 200~Myr, leaving
325~evolved models to analyse.
\par The number of particles is fixed to $10^5$ for the NS+disc components
and $10^5$ for the DH. The proportion of particles in the NS and in the disc
is such that the particle masses are exactly the same for both components, and
may therefore change from one simulation to another: e.g. 30262 NS particles
and 69738 disc particles for m00.
\subsection{$N$-body code}
The initial models are integrated with the Particle-Mesh code described in
Pfenniger \& Friedli (\cite{PF}).
\par The potential is computed on a cylindrical polar grid using the fast
Fourier transform technique in the $\phi$ and $z$ dimensions, where the cells
are equally spaced. The radial spacing of the cells is logarithmic with a
linear core to avoid an accumulation point at the centre. The short range
forces are softened by a variable homogeneous ellipsoidal kernel with
semi-axes set to 1.1 times the respective cell dimensions. The adopted grid
has $25(R)\times 24(\phi)\times 201(z)$ cells and extends up to 50~kpc in $R$
and $\pm 20$~kpc in $z$, corresponding to a radial resolution of 40~pc at the
centre and about 1.8~kpc at $R_{\circ}$.
\par The orbits are integrated using the standard leap-frog algorithm with a
time step of 0.1~Myr, representing a fraction of the crossing time of the
central grid mesh in the steep NS potential.
\subsection{Model evolution}
%
\begin{figure*}
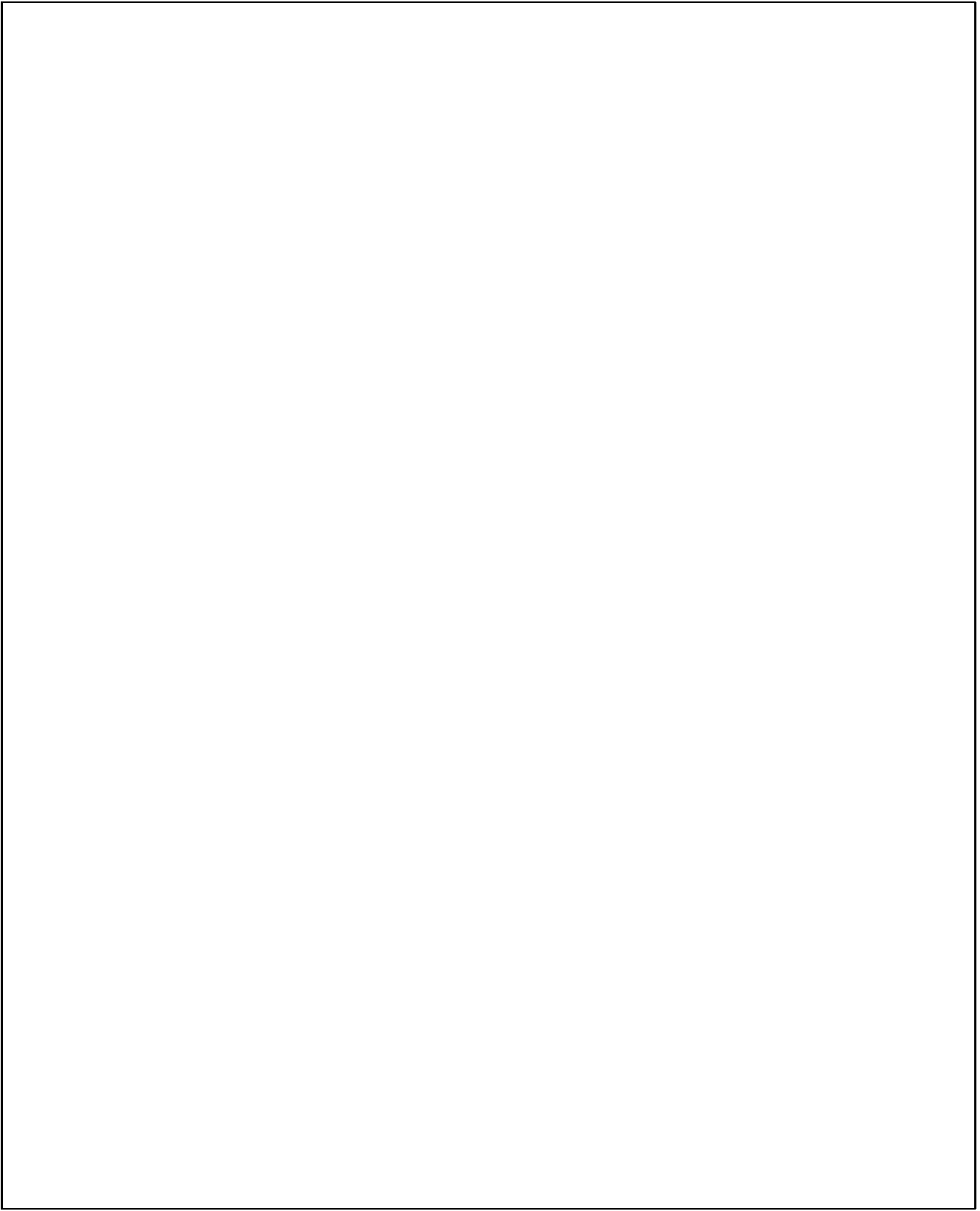

\picplace{22.3cm}
\caption[]{Face-on surface density evolution in all runs, including 1/3 of all
investigated models. The distances are in initial units and the contours are
spaced with one magnitude interval. Rotation is clockwise. Note the induced
asymmetries in run~m12}
\label{evol}
\end{figure*}
Figure~\ref{evol} illustrates the evolution of our 13 simulations. They
all lead to the formation of a bar, but not always on the same time scale.
Models with higher initial values of Toomre's axisymmetric stability parameter
$Q$ in the region of rising circular velocity need more time to develop the
bar, in agreement with the work of Athanassoula \& Sellwood (\cite{AS}). This
is the case for example when comparing the simulations~m07 and m08: the former
starts with a much colder disc, i.e a lower value of $Q$, and indeed forms the
bar more quickly.
\par The size of the bars clearly varies from one simulation to the other.
However, one must keep in mind here that the evolved models can always be
separately rescaled to look more similar than they do in initial units, hence
complicating an objective comparison. All bars finally flatten the surrounding
radial profile of the disc, the effect being particularly marked in
simulation~m03.
\par Some simulations obviously pass through a double bar phase, like in
m04t4400. The simulations also develop transient spiral arms, especially
strong during the time of bar formation. Such structures would be hard to
achieve by other means than the $N$-body technique.
\par
\begin{figure}[t]
\centerline{\psfig{figure=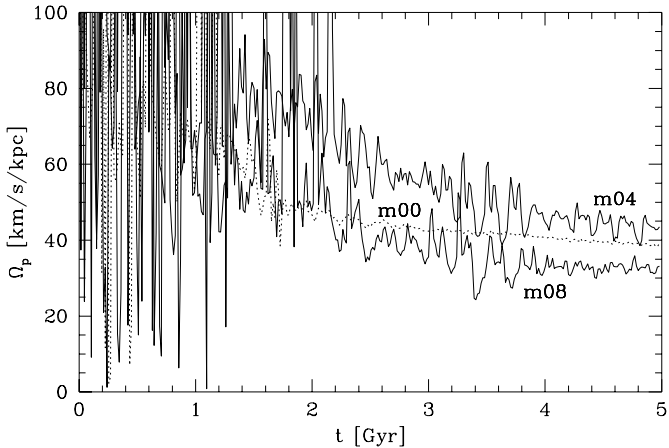,width=8.8cm}}
\caption[]{Pattern speed $\Omega_{\rm P}(t)$ of the bar in simulations~m00,
m04 and m08 (initial units). The pattern speed in the other simulations
roughly fall within the gap delimited by the m04 and m08 solid lines}
\label{omp}
\end{figure}

 Figure~\ref{omp} shows the bar pattern speed $\Omega_{\rm P} \equiv
\frac{d\vartheta}{dt}(t)$ as a function of time in three simulations.
The azimuthal angle $\vartheta$ of the bar major axis (in the inertial frame)
is calculated by diagonalising the $I_{xx}$, $I_{yy}$ and $I_{xy}$ components
of the moment of inertia tensor of the NS+disc particles inside $R<R_{\rm e}$,
where $R_{\rm e}$ is the radius extremising the ratio of the inferred diagonal
components. The noise level in the $\Omega_{\rm P}(t)$ curves depends on the
bar strength and on the presence of multiple rotating patterns.
\par After the first bar rotations, where the bar may slow down by up to
15 km/s/kpc/Gyr, $\Omega_{\rm P}(t)$ is in general constant or slowly
decreasing with a rate of a few km/s/kpc/Gyr. The lower rotation of the DH in
simulation~m11 does not influence the pattern speed of the bar:
$\Omega_{\rm P}(t)$ in m11 is almost the same as in m00. Simulation~m12 has a
faster $\Omega_{\rm P}(t)$ decline than m00, probably related to its noisier
gravitational potential.
\par The presence of a dissipative gas component may alter the bar evolution
by speeding up its angular velocity and causing its dissolution (Friedli \&
Benz \cite{FR}).  
%
\section{Location of the observer}
\label{adj}
The Diffuse Infrared Background Experiment (DIRBE) onboard the COBE satellite
has mapped the full sky in ten different bands, ranging from $1.25 \mu$m to
$240 \mu$m, with a resolution of $0\fdg 7 \times 0\fdg 7$ and a field spacing
of approximately $0\fdg 32$. The maps at 1.25 ($J$-band), 2.2 ($K$-band),
3.5 and $4.9\mu$m, dominated by integrated stellar light, clearly show an
asymmetric boxy/peanut shaped bulge betraying the underlying bar (Weiland et
al. \cite{WA}).
\par The $K$-band map offers a good compromise between maximum stellar
emissivity, low extinction by dust and small relative contribution of zodiacal
light. If a constant mass-to-light ratio is assumed throughout the Galaxy, it
should therefore provide a reliable tracer of the integrated mass distribution
and thus put severe constraints on dynamical models of the Milky Way.
\par This section presents a technique to find the view point in an $N$-body
galaxy from where the simulated panorama most ressembles a fixed map, and
applies the method to our models and a deredened version of the COBE $K$-band
data.
\subsection{Reduction of the COBE $K$-band map}
Even if extinction in the infrared is much less than in the optical, it is
still not negligeable (about 1.5~magnitude in $K$ towards the Galactic centre).
\par Thus, following Arendt et al. (\cite{AB}), the raw $K$-band map has been
corrected for {\it foreground} dust extinction by first building a $J\!-\!K$
color excess map under the assumption of constant intrinsic color, taken as
the average color in the region $l<30\degr$ and $10\degr \leq b \leq
15\degr$, and then transforming it into a $K$-band optical depth map using
the redening law of Rieke \& Lebofsky (\cite{RL}). Such a correction however
is not valid for $|b|\la 3\degr$ where a significant fraction of the
integrated light is emitted along the dust screen. Hence this low latitude
region must be excluded in the fitting procedure.
\par Finally, the dust subtracted map has been symmetriesed in $b$ and
converted from its COBE Quadrilateralised Spherical Cube representation to a
cartesian grid map in Galactic coordinates with $\Delta l=\Delta b=1\degr$
pixel size, as shown in Fig.~\ref{limit}. This resolution ensures sufficient
particle statistics per pixel when computing model maps without bluring too
much the data.
\subsection{Fitting method}
\label{met}
The position of the observer in the models is specified by its distance
$\tilde{R}$ in initial units from the ``Galactic'' centre and the scale
free-angle $\varphi$ between the line joining himself to the centre and the
major axis of the bar, with positive $\varphi$ when the bar is leading the
observer. We assume that the observer lies in the plane of symmetry,  i.e.
$z_{\circ}=0$, and that the visible components have a constant
mass-to-light ratio $\Upsilon_K$ in the $K$-band. Hence model maps of
integrated light will depend on $\tilde{R}$, $\varphi$ and $\Upsilon_K$.
\par The flux per unit solid angle in a given pixel $i$ is estimated from
Monte Carlo integration:
\begin{equation}
F_i(\tilde{R},\varphi,\Upsilon_K)=(\Upsilon_K \Delta\Omega)^{-1}
\sum_{k=1}^{N_i}\frac{m}{D_k^2+\varepsilon^2},
\label{F}
\end{equation}
where the sum ranges over all NS+disc particles inside pixel $i$, $N_i$ is the
total number of such particles, $m$ the mass per particle (identical for both
visible components), $D_k$~the distance of the $k^{\rm th}$~particle relative
to the observer, $\varepsilon$ a softening parameter to reduce the statistical
noise induced by the closest particles ($\varepsilon^2=0.1$), and
$\Delta\Omega$ the solid angle sustained by the pixel. If $b_1$ and
$b_2=b_1+\Delta b$ are the pixel limits in Galactic latitude, then:
\begin{equation}
\Delta\Omega = \Delta l(\sin{b_2}-\sin{b_1}).
\end{equation}
\begin{figure}[t]
\centerline{\psfig{figure=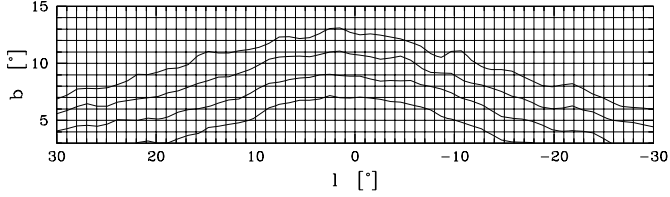,width=8.8cm}}
\caption[]{Grid used for the data interpolation and the model flux calculation
in the COBE-map fitting. The solid lines roughly delimit the 100, 200, 300
and 400 pixels which in the models have the best particle statistics}
\label{limit}
\end{figure}
\par The three parameters $\tilde{R}$, $\varphi$ and $\Upsilon_K$ are adjusted
to the COBE/DIRBE $K$-band data by minimising the mean quadratic relative
residual between model and observed fluxes {\it corrected} for statistical
noise, i.e. (see Appendix~B):
\begin{equation}
{\cal R}^2(\tilde{R},\varphi,\Upsilon_K)=(\chi^2-\nu)
\left[\sum_{i=1}^{N_{\rm pix}}\frac{{F_i^{\circ}}^2}{\sigma_i^2}\right]^{-1},
\label{calR}
\end{equation}
where
\begin{equation}
\hspace{1cm}
\chi^2=\sum_{i=1}^{N_{\rm pix}}\frac{(F_i-F_i^{\circ})^2}{\sigma_i^2} \;,
\label{chi2}
\end{equation}
$N_{\rm pix}$ is the number of pixel, $\nu=N_{\rm pix}-3$ the number of degree
of freedom, $F_i^{\circ}$ the observed flux in pixel $i$, and $\sigma_i^2$ the
variance of $F_i$, whose best estimate reduces to (see Appendix~C):
\begin{equation}
\sigma_i^2=(\Upsilon_K \Delta\Omega)^{-2}\sum_{k=1}^{N_i}\left(
\frac{m}{D_k^2+\varepsilon^2}\right)^2.
\label{var}
\end{equation}
The data supply much more accurate fluxes than the models, so that their errors
may be neglected.
\par The standard $\chi^2$~minimisation method was abandoned
because the $\chi^2$ strongly depends on the precision of the $F_i$'s.
In particular, the $\chi^2$ value will systematically increase with the number
of particles available to compute the model maps (except for perfect models).
Thus the fitted view points are biased towards regions where the model maps are
noisier, and the relative quality of different models cannot be judged from
this indicator alone. Furthermore, the fact that the $\sigma_i^2$'s are only
approximations of the true variances causes an overestimation of the $\chi^2$
increasing with the uncertainties on the variances, which biases the solutions
in the opposite way. This bias persists even in ${\cal R}^2$~adjustments
since ${\cal R}^2$ depends explicitly on $\chi^2$. To reduce its effect, the
$\sigma_i^2$'s have been averaged over the neighbouring pixels.
\par Only the bulge region $|l|<30\degr$ and $3\degr<b<15\degr$
enclosing the bar and with reliable dust correction, representing 720 pixels,
is included in the fits. Moreover, to exclude pixels with low number of
particles and to check the consistency of the fitted parameters with respect
to the selected $l\!-\!b$ sub-region, the pixels are sorted by decreasing
$N_i$ and only the most populated are retained, with $N_{\rm pix}$ between 100
and 400. The typical $l\!-\!b$ regions selected this way are indicated in
Fig.~\ref{limit}.
\par For each model, ${\cal R}^2$ is computed on a 2D grid in $\tilde{R}$ and
$\varphi$ with resolution $\Delta \tilde{R}=0.1$~kpc and $\Delta \varphi=
1\degr$. The parameter $\Upsilon_K$ does not require an extra dimension,
since for fixed values of $\tilde{R}$ and $\varphi$ the optimum $\Upsilon_K$,
which minimises ${\cal R}^2$, can be calculated analytically as:
\begin{equation}
\Upsilon_{K,{\rm min}}(\tilde{R},\varphi)=
\left(\sum_{i=1}^{N_{\rm pix}}\frac{F_i'^2}{\sigma_i'^2}-\nu\right)
\left(\sum_{i=1}^{N_{\rm pix}}\frac{F_i' F_i^{\circ}}{\sigma_i'^2}\right)^{-1},
\label{min}
\end{equation}
where the primes indicate that the $\Upsilon_K$ factor in Eqs.~(\ref{F}) and
(\ref{var}) is removed. The ${\cal R}^2$-grid is then smoothed by averaging
${\cal R}^2$ over the $3\times 3$ nearest grid points and the resulting
minimum defines our best fit position of the observer.
\subsection{Tests}
The method has been tested with a simple mass model consisting in a triaxial
multi-normal bulge for the bar and an axisymmetric Miyamoto-Nagai (\cite{MN})
disc:
\begin{eqnarray}
\rho_{\rm test}(x,y,z) & = & \frac{M_1}{\sqrt{8\pi^3}\sigma_x\sigma_y\sigma_z}
     \exp{\left[-\frac{1}{2}\left(\frac{x^2}{\sigma_x^2}+\frac{y^2}{\sigma_y^2}
     +\frac{z^2}{\sigma_z^2}\right)\right]} \nonumber \\
&   & \hspace{-1cm}+\frac{b_{\rm M}^2M_2}{4\pi}\left[\frac{a_{\rm M}R^2
      +(a_{\rm M}+3z_{\rm b})(a_{\rm M}+z_{\rm b})^2}{(R^2+(a_{\rm M}
      +z_{\rm b})^2)^{5/2}z_{\rm b}^3}\right],
\label{rhotest}
\end{eqnarray}
where $R^2=x^2+y^2$, $z_{\rm b}^2=z^2+b_{\rm M}^2$, and with $\sigma_x=1.5$,
$\sigma_y=0.8$, $\sigma_z=0.4$, $a_{\rm M}=5$ and $b_{\rm M}=0.3$. The
Miyamoto-Nagai component is truncated at $R=15$ and its mass within this
radius is $3M_1$.
\par
\begin{figure}
\centerline{\psfig{figure=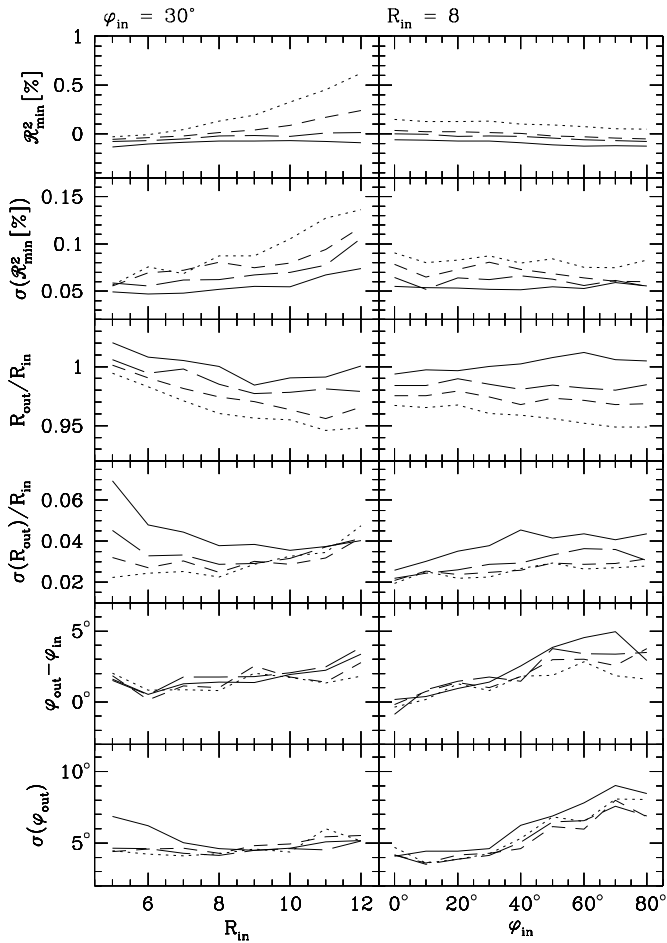,width=8.8cm}}
\caption[]{Test results of the ${\cal R}^2$~minimisation technique.
$R_{\rm in}$ and $\varphi_{\rm in}$ are the correct input coordinates of the
observer and $R_{\rm out}$ and $\varphi_{\rm out}$ the corresponding adjusted
output parameters. The left and right columns are resp. for variable
$R_{\rm in}$ and fixed $\varphi_{\rm in}$, and for fixed $R_{\rm in}$ and
variable $\varphi_{\rm in}$. From top to bottom, based on 100 experiments for
each input position: the averages and standard deviations of ${\cal R}^2$ at
the minimum, $R_{\rm out}/R_{\rm in}$ and $\varphi_{\rm out}$. The full,
long-dashed, short-dashed and dotted lines stand resp. for $N_{\rm pix}=100$,
200, 300 and 400}
\label{test}
\end{figure}

 COBE-like maps were computed for several {\it input} observing points
$(R_{\rm in},\varphi_{\rm in})$ in the test model and 100 distinct discrete
realisations of this model with $10^5$ particles, as in the NS+disc components
of the dynamical models, were fitted to these maps by ${\cal R}^2$
minimisation, yielding the {\it output} observing points $(R_{\rm out},
\varphi_{\rm out})$. The statistics of the results is shown in
Fig.~\ref{test}.
\par For $\varphi_{\rm in}\la 40\degr$ and $R_{\rm in}\la 10$, the systematic
and statistical errors on $\varphi_{\rm out}$ are at most $+2\degr$ and
$5\degr$ (except at small $R_{\rm in}$ for $N_{\rm pix}=100$).
For $N_{\rm pix}\geq 200$, the statistical error on $R_{\rm out}$ is less than
4\%, and $R_{\rm out}$ is underestimated with a systematic error increasing
with $N_{\rm pix}$, which is intimately connected with the $\chi^2$ bias
already mentioned in~Sect.~\ref{met}: the additional pixels in the low flux
region have the most uncertain $\sigma_i^2$'s and thus artificially enhance
the value of ${\cal R}^2$, as indicated in the top plots of Fig.~\ref{test}.
\begin{figure}[t]
\centerline{\psfig{figure=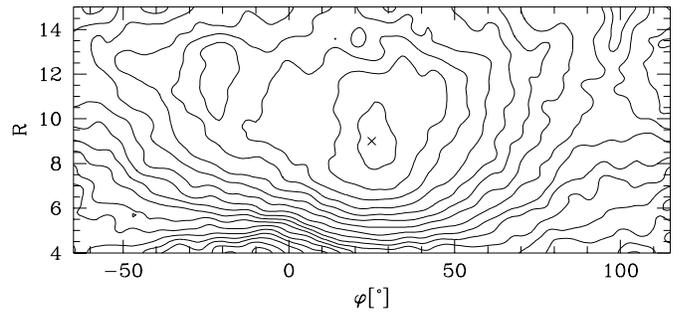,width=8.8cm}}
\caption[]{${\cal R}^2$ solution for the location of the observer in model
m08t3200, for $N_{\rm pix}=300$. The ${\cal R}^2$ contours are spaced by 0.5\%
and the cross indicates the position of the minimum. The parameter
$\Upsilon_K$ has been substituted using Eq.~(\ref{min})}
\label{Q2}
\end{figure}
\subsection{Application}
The ${\cal R}^2$ method has been applied to all dynamical \mbox{models}, noting
$\varphi_{\circ}$ and $\tilde{R}_{\circ}$ the best fit parameters of the
observer and ${\cal R}_{\rm min}^2$ the value of ${\cal R}^2$ at the minimum.
An example of ${\cal R}^2$~topography is given in Fig.~\ref{Q2}.
\par As a compromise between maximum exploitation of the data and minimum
systematic errors on the fitted view point parameters, we have privileged
the case $N_{\rm pix}=300$ and sorted the models by increasing ${\cal R}_{\rm
min}^2 (300)$. This model sequence was then truncated to include at least two
models of each simulation, resulting in a sample of 68 models with
${\cal R}_{\rm min}^2(300) \leq 0.756$\% (hereafter our ``A''~sample).
\par
\begin{figure*}
\centerline{\psfig{figure=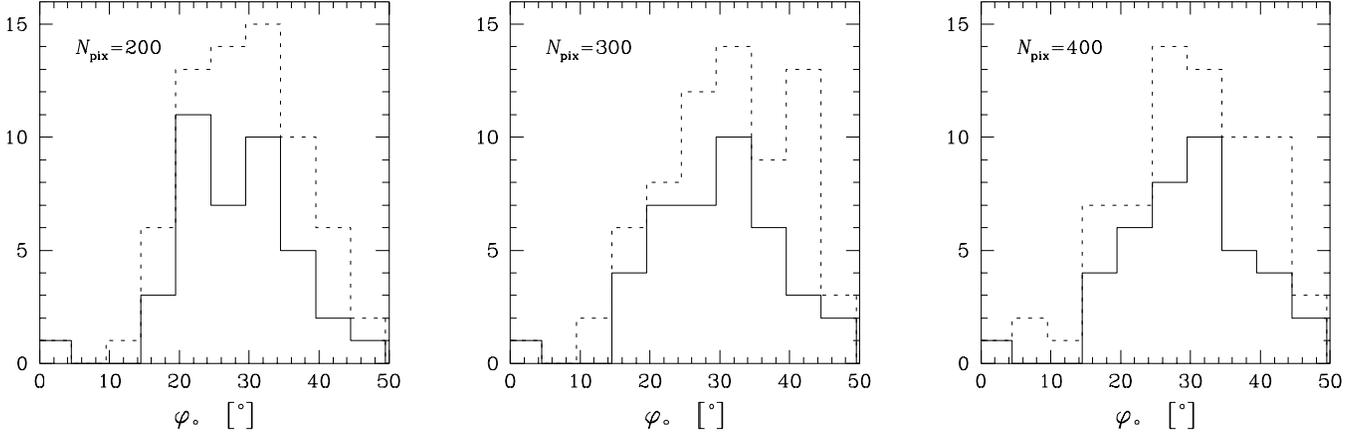,width=18cm}}
\caption[]{Distribution of the best-fit position angle $\varphi_{\circ}$ of
the bar in the dynamical models. Dashed line: including all 68 models of the
A~sample; full line: including only the 40 lowest ${\cal R}_{\rm min}^2$
models of this sample}
\label{hist}
\end{figure*}

 Figure~\ref{hist} shows the distribution of the adjusted angles
$\varphi_{\circ}$ for the models belonging to this sample as a function
of $N_{\rm pix}$. Taking into account the uncertainties of the individual
angles, due to the moment of inertia method for determining the absolute
angular position of the bar major axis and to the intrinsic standard deviation
of the ${\cal R}^2$ method, the 40 best models support an angle of
$\varphi_{\circ}=28\degr \pm 7\degr$ for the Galactic bar, with a possible
overestimation by $2\degr$ according to the tests.
\par This result is consistent with the value of $20\degr\!-30\!\degr$ found
recently by the OGLE team from the color magnitude diagram of red clump giants
in several fields across the Galactic bulge (Stanek et al. \cite{SU}) and with
the upper limit of $30\degr$ set by the MACHO microlensing constraints
(Gyuk~\cite{GG}), and confirms the privileged $25\degr$ obtained by the
deprojection of the properly dust subtracted $L$-band COBE map (Binney et al.
\cite{BGS}; Bissantz et al. \cite{BE}).
\par The best models regarding the COBE constraints come from simulation~m08,
started with the thickest and hottest disc. Indeed five of the six lowest
residual models are from this simulation, with ${\cal R}_{\rm min}^2(300)
\approx 0.3\!-\!0.4$\%. For comparison, the initial m00~model has ${\cal R}^2
\geq 2.4$\% everywhere. Nevertheless, we note that model m06t4600 gives the
best COBE-data match within the entire $|l|<90\degr$ region. This model has a
low disc scale height, close to the 204 pc derived by Kent et al. (\cite{KDF})
from the Spacelab IR Telescope data with the assuption of constant $h_z$.
\par Table~\ref{cobe} lists the results for a selection of A sample models
(hereafter our ``B''~sample) which, in addition to the COBE constraints, also
best reproduce observations in the Solar neighbourhood (see Sect.~\ref{sol}).
The simulation~m00 provide no acceptable model to enter this subsample.
\par The distance scale $S_R$ of the models now directly follows by setting
the best-fit Galactocentric distance of the observer for $N_{\rm pix}=300$ to
$R_{\circ}=8$~kpc. Figure~\ref{qq} presents the NS+disc face-on configurations
of some B~sample models after such rescaling and the confrontation of these
models to the COBE data. The B~sample models have a face-on surface density
axis ratio $b/a=0.5\pm 0.1$ for the contours crossing the bar major axis at
$R=2\!-\!3$~kpc.
\begin{table*}[htbp]
\centering
\caption[]{Location of the observer in the final selection of models
(B~sample), as constrained by the COBE/DIRBE deredened $K$-band map. The
models are sorted by increasing ${\cal R}_{\rm min}^2(N_{\rm pix}=300)$. The
last two columns give the adopted distance ($S_R$) and velocity ($S_V$) scales
based on the $N_{\rm pix}=300$ solutions}
\begin{tabular}{crclrclrclrc} \hline
   & \multicolumn{3}{c}{$N_{\rm pix}=200$  } &
     \multicolumn{3}{c}{$N_{\rm pix}=300$~~} &
     \multicolumn{3}{c}{$N_{\rm pix}=400$~~} \\
Model & ~~~~~~$\tilde{R}_{\circ}$ & $\varphi_{\circ}[\degr]$ &
\multicolumn{1}{l}{\hspace{-.15cm}${\cal R}_{\rm min}^2$ [\%]}
      &    ~~~$\tilde{R}_{\circ}$ & $\varphi_{\circ}[\degr]$ &
\multicolumn{1}{l}{\hspace{-.15cm}${\cal R}_{\rm min}^2$ [\%]}
      &    ~~~$\tilde{R}_{\circ}$ & $\varphi_{\circ}[\degr]$ &
\multicolumn{1}{l}{\hspace{-.15cm}${\cal R}_{\rm min}^2$ [\%]}
      &    ~~~~$S_R$~ & $S_V$ \\ \hline
m08t4000 &  {\it 9.3} & {\it 31} & {\it 0.228} &  9.5 & 33 & 0.342 &
{\it 9.4} & {\it 38} & {\it 0.359} & 0.842 & 0.900 \\
m12t2000 &  {\it 8.2} & {\it 23} & {\it 0.319} &  8.2 & 23 & 0.409 &
{\it 8.2} & {\it 23} & {\it 0.337} & 0.976 & 0.939 \\
m06t4600 &  {\it 9.6} & {\it 26} & {\it 0.388} &  9.6 & 26 & 0.432 &
{\it 9.6} & {\it 26} & {\it 0.619} & 0.833 & 1.120 \\
m11t2000 &  {\it 8.1} & {\it 27} & {\it 0.365} &  9.0 & 34 & 0.467 &
{\it 8.2} & {\it 28} & {\it 0.526} & 0.889 & 0.964 \\
m08t3200 &  {\it 8.6} & {\it 24} & {\it 0.424} &  9.0 & 25 & 0.472 &
{\it 8.7} & {\it 25} & {\it 0.515} & 0.889 & 0.977 \\
m09t1600 &  {\it 8.5} & {\it 26} & {\it 0.341} &  8.5 & 26 & 0.586 &
{\it 8.5} & {\it 26} & {\it 0.828} & 0.941 & 0.991 \\
m02t2000 &  {\it 8.1} & {\it 21} & {\it 0.547} &  8.1 & 21 & 0.628 &
{\it 8.0} & {\it 23} & {\it 0.672} & 0.988 & 0.946 \\
m03t1000 &  {\it 7.5} & {\it 22} & {\it 0.599} &  9.5 & 29 & 0.641 &
{\it 8.0} & {\it 21} & {\it 0.728} & 0.842 & 1.056 \\
m12t1600 &  {\it 7.3} & {\it 20} & {\it 0.557} &  7.3 & 19 & 0.677 &
{\it 7.5} & {\it 19} & {\it 0.780} & 1.096 & 1.001 \\
m10t1400 &  {\it 7.4} & {\it 19} & {\it 0.580} &  8.1 & 22 & 0.709 &
{\it 7.4} & {\it 15} & {\it 0.730} & 0.988 & 0.916 \\
m04t3000 &  {\it 7.9} & {\it 18} & {\it 0.615} &  7.9 & 32 & 0.738 &
{\it 7.7} & {\it 25} & {\it 0.784} & 1.013 & 0.852 \\
m06t4800 &  {\it 8.0} & {\it 26} & {\it 0.675} &  8.0 & 25 & 0.747 &
{\it 9.3} & {\it ~9} & {\it 0.650} & 1.000 & 1.137 \\
                                      \hline
\end{tabular}
\label{cobe}
\end{table*}
\begin{table*}[htbp]
\centering
\caption[]{Several absolute properties of the rescaled B~sample models.
$\chi^2_{\rm loc}$: mean square residual between model and observed local
properties; $\sigma^D,\sigma^S$: local disc and NS velocity dispersions~[km/s];
$\overline{v_{\phi}}^S$: local NS rotation velocity~[km/s]; $h_R,h_z$: disc
scale length~[kpc] and local scale height~[pc]; $\Sigma_{\circ}^D,
\rho_{\circ}^S$: local disc surface density~[$M_{\odot}/$pc$^2$] and NS volume
density~[$10^{-3}M_{\odot}/$pc$^3$]; $V_{\circ}$:~local circular
velocity~[km/s]; $\Omega_{\rm P},R_{\rm L}$: pattern speed of the
bar~[km/s/kpc] and corotation radius~[kpc]; $\varphi_{\circ}$: bar inclination
angle~[$\degr$]; $\Upsilon_K$: mass-to-K band luminosity
ratio~[$M_{\odot}/L_{K,\odot}$]; $\tau_0$: microlensing optical depth towards
Baade's Window~[$10^{-6}$]. Values in brackets include the DH lenses. The
boldfaced models reasonably agree with most of the considered observations}
\begin{tabular}{crcccrccccr} \hline
&&&&&&&&& \vspace{-.33cm} \\
Model & $\chi_{loc}^2$ & $\sigma_{RR}^D \sigma_{\phi\phi}^D \sigma_{zz}^D$ &
$\sigma_{RR}^S\;\,\sigma_{\phi\phi}^S\;\,\sigma_{zz}^S\;\,
\overline{v_{\phi}}^S\!\!$ & $h_R$~~~$h_z$ & $\Sigma_{\circ}^D\;\,
\rho_{\circ}^S$& $V_{\circ}$ & $\Omega_{\rm P}$~$R_{\rm L}$ &
$\varphi_{\circ}$ & $\Upsilon_K$ & $\tau_0$~~~~~~~ \\ \hline
 m08t4000 &  2.98 & 43~~ 28~~ 21 & 106~~ 117~~  78~~ 51 &
  2.7~~ 363 & 47~~ 0.9 & 195 & 36~~ 5.5 & 33 & 0.60 & 1.53~(1.75) \\
 m12t2000 &  1.91 & 44~~ 27~~ 19 & 118~~ 107~~  89~~ 63 &
  4.9~~ 339 & 49~~ 1.2 & 200 & 44~~ 4.5 & 23 & 0.68 & 1.70~(2.12) \\
 m06t4600 &  5.59 & 36~~ 26~~ 17 & 135~~ 129~~  95~~ 41 &
  4.0~~ 232 & 58~~ 1.3 & 230 & 54~~ 3.9 & 26 & 0.59 & 1.45~(1.61) \\
 m11t2000 &  1.88 & 53~~ 26~~ 20 & 117~~ 116~~  83~~ 61 &
  2.7~~ 287 & 40~~ 1.2 & 210 & 52~~ 4.0 & 34 & 0.57 & 1.38~(1.48) \\
{\bf m08t3200} &  3.88 & 42~~ 25~~ 22 & 111~~ 124~~  81~~ 58 &
  2.7~~ 379 & 44~~ 1.1 & 214 & 44~~ 4.8 & 25 & 0.72 & 1.80~(1.97) \\
 m09t1600 &  3.61 & 38~~ 24~~ 18 & 124~~ 121~~  84~~ 61 &
  4.4~~ 259 & 50~~ 1.2 & 212 & 55~~ 3.9 & 26 & 0.73 & 2.19~(2.33) \\
 m02t2000 &  1.64 & 51~~ 29~~ 20 & 115~~ 117~~  94~~ 60 &
  3.1~~ 290 & 43~~ 1.1 & 210 & 56~~ 3.9 & 21 & 0.73 & 1.97~(2.07) \\
 m03t1000 &  2.47 & 41~~ 30~~ 18 & 129~~ 125~~  93~~ 44 &
  4.9~~ 271 & 45~~ 0.6 & 216 & 58~~ 3.9 & 29 & 0.53 & 1.54~(1.89) \\
 m12t1600 &  3.43 & 50~~ 30~~ 25 & 118~~ 115~~  96~~ 76 &
  3.9~~ 328 & 70~~ 1.7 & 217 & 48~~ 4.6 & 19 & 0.96 & 3.00~(3.13) \\
 m10t1400 &  1.82 & 54~~ 28~~ 18 & 119~~ 111~~  91~~ 64 &
  5.1~~ 295 & 47~~ 1.2 & 198 & 44~~ 4.9 & 22 & 0.70 & 1.97~(2.18) \\
{\bf m04t3000} &  3.40 & 36~~ 24~~ 19 & 111~~ 106~~  84~~ 50 &
  3.0~~ 285 & 39~~ 1.5 & 200 & 48~~ 4.3 & 32 & 0.72 & 2.03~(2.11) \\
m06t4800 &  2.39 & 47~~ 27~~ 23 & 133~~ 133\hspace{.1cm}104\hspace{.05cm} 50 &
  4.6~~ 278 & 58~~ 1.7 & 229 & 45~~ 4.6 & 25 & 0.79 & 2.28~(2.46) \\
                                                                        \hline
\end{tabular}
\label{out}
\end{table*}
\begin{figure*}
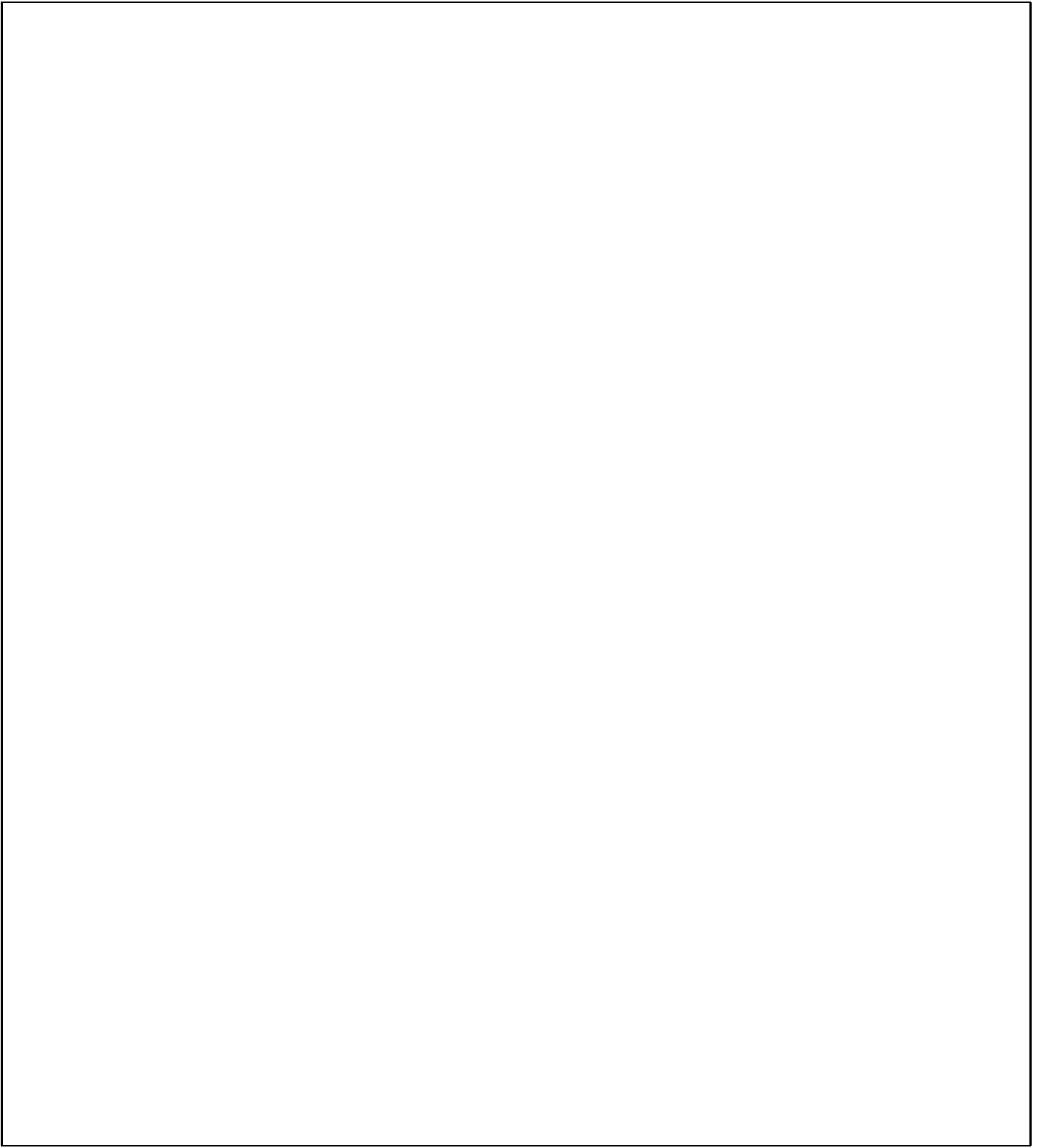

\picplace{20cm}
\caption[]{Comparison of a selection of rescaled models with observations in
the bar region. Left: face-on surface density of the visible components; The
symbol $\odot$ indicates the location of the observer and the crosses the
positions of the Lagrangian points $L_1$ and $L_2$. Rotation is clockwise.
Middle: COBE/DIRBE deredened K-band contours (full lines) and corresponding
model contours (dashed lines). The $b$ scale is dilated relative to $l$, hence
amplifying the deviations. The region below the horizontal line was excluded
in the adjustments because of unreliable extinction correction. The spacing
between the contours is 0.5 magnitude in both frames. Right: longitude-velocity
diagram of the Galactic HI averaged over $|b|<1.25\degr$ and, superposed,
the traces of non-self intersecting model $x_1$-orbits (dashed lines); the
innermost trace represents the cusped $x_1$ orbit. The model m12t2000 has been
symmetrised in all frames. The HI data were kindly provided by H. Liszt and
refer to Burton \& Liszt (\cite{BL})}
\label{qq}
\end{figure*}
\addtocounter{figure}{-1}
\begin{figure*}
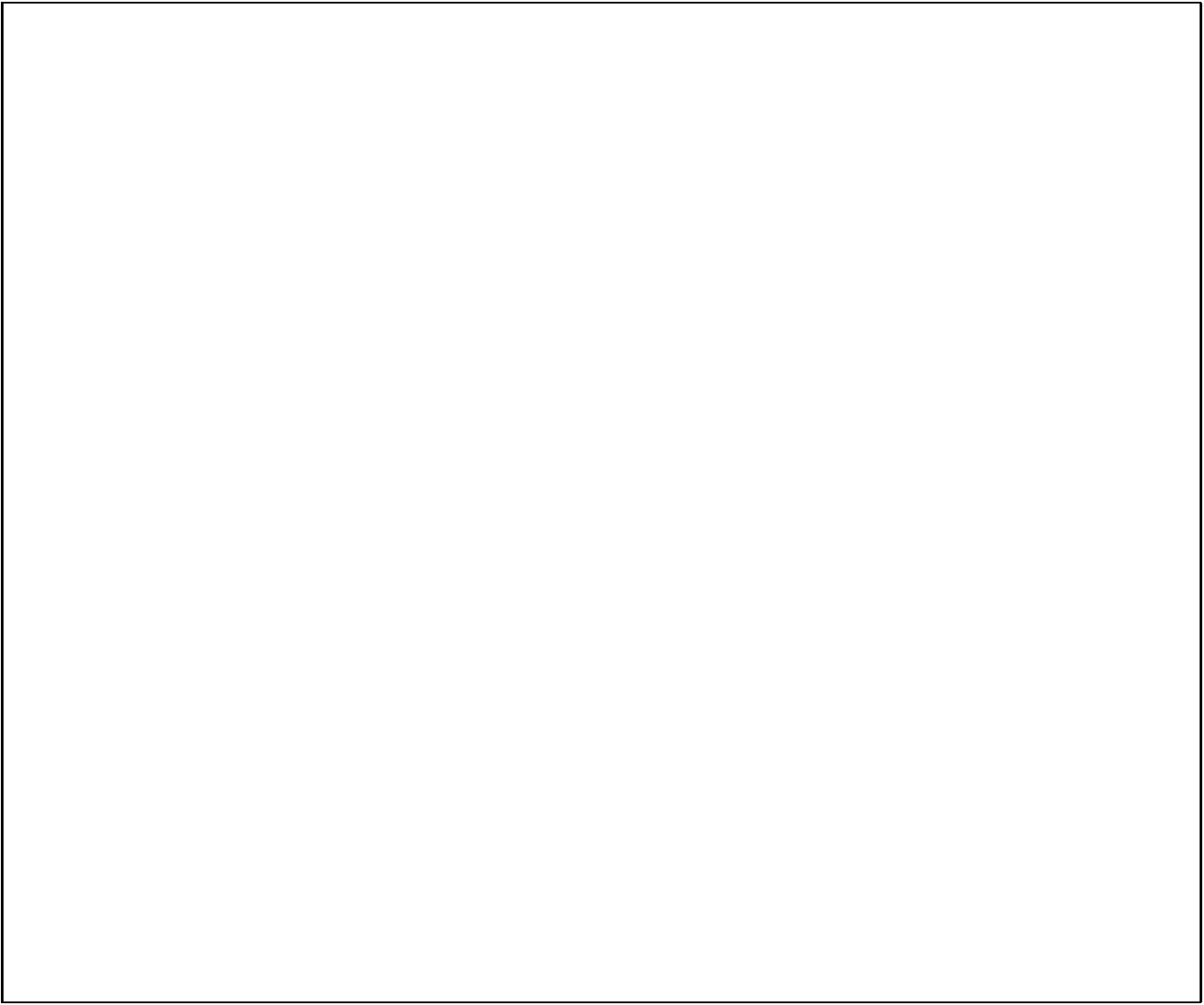

\picplace{15cm}
\caption[]{continued}
\label{qq2}
\end{figure*}
%
\section{Model properties and discussion}
\label{disc}
Given the distance scale, there remains only one scale left in the models since
the constraint of virial equilibrium links together the distance, velocity
and mass scales. We choose here to fix the velocity scale by requiring that
the line-of-sight velocity dispersion of the combined NS+disc components
towards Baade's Window $(l,b)=(0\fdg 9,-3\fdg 9)$ equals $113\pm 6$~km/s, as
observed for late-M giants (Sharples et al. \cite{SW}), leaving about 10\%
uncertainty on the absolute mass scale. The resulting velocity scales of the
B~sample models are listed in Table~\ref{cobe}.
\par The complete scaling of the models allows now to derive several of their
absolute properties, which are presented and discussed in this section.
\subsection{Local properties}
\label{sol}
The main local properties of the models, i.e. the properties measured at the
location of the observer, are given in Table~\ref{out}. These include
the velocity dispersion of the disc ($\sigma_{RR}^D,\sigma_{\phi\phi}^D,
\sigma_{zz}^D$) and of the NS ($\sigma_{RR}^S,\sigma_{\phi\phi}^S
\sigma_{zz}^S$), the NS mean rotation velocity ($\overline{v_{\phi}}^S$), the
disc scale length ($h_R$) and scale height ($h_z$), the disc surface density
($\Sigma_{\circ}^D$), the NS volume density ($\rho_{\circ}^S$) and the circular
velocity ($V_{\circ}$).
\par All quantities, except $h_R$ and $V_{\circ}$, are based on the
particles inside a sphere (for the NS) or a vertical cylinder (for the disc)
centred on the Sun's position, and hence are not azimuthal averages. The disc
scale length is an effective exponential scale length derived from all disc
particles within the annulus $0.7<R/R_{\circ}<1.2$. The circular velocity
rests on the azimuthally symmetrised potential.
\par To quantify the ability of the models to reproduce local observations, 
Table~\ref{out} also gives the mean $\chi^2$~residual between the local model
properties, from $\sigma_{RR}^D$ to $\Sigma_{\circ}^D$ and excluding $h_R$,
and the corresponding observed values mentioned in Table~\ref{loc}, weighting
the contribution of the three disc velocity dispersion components by $1/3$ and
the four NS velocity moments by $1/4$. Models from simulation~m07 have very
high $\chi_{\rm loc}^2$ due to their small disc scale height, as well as
unrealistic mass-to-light ratios and too low microlensing optical depths in
Baade's Window (see Sect.~\ref{mtl} and \ref{opt}), and were therefore
rejected.
\par Many models have disc kinematics consistent with observations, and in
particular correct ratios between the components of the velocity dispersion.
There appears a clear correlation between the vertical velocity dispersion of
the disc and its scale height, as expected from self-gravitating isothermal
sheets, although not directly depending on the disc surface density. A proper
analysis of the disc vertical dynamics would of course require simulations
with higher $z$-resolution. The disc velocity dispersion in model m06t4600 is
lower than in m06t4800 because the observer is located farther from the centre
(in initial units) and the velocity dispersion decreases outwards.
\par The velocity dispersion of the NS components is in general similar to
that of the local Galactic halo stars, except that its radial anisotropy is
insufficient. This difference is probably linked to the initial model
truncation at $R_{\rm c}=38$~kpc, which limits the amount of radial anisotropy
inside the models, and the flatness ($e=0.5$) of the outer mass distribution,
which requires substantial support by azimuthal motion, either in systematic
or random form. It is however worthwhile to mention that the local subdwarf
kinematics in Table~\ref{loc} may be biased by selection effects. For example,
kinematically selected samples have much larger radial anisotropy than samples
selected by metallicity criterions, the velocity dispersion in the latter
being even consistent with meridional isotropy (Norris \cite{N}). Moreover,
the kinematics of the globular cluster system shows no significant deviation
from isotropy.
\par All models favour $V_{\circ}<220$~km/s except those from simulation~m06.
These models have in fact increasing rotation curve from $R_{\circ}/2$ to
$R_{\circ}$, so that our velocity scaling mainly based on the inner region
leads to larger circular velocities at $R_{\circ}$.
\subsection{Corotation and terminal velocities from $x_1$ orbits}
\begin{figure}[t]
\centerline{\psfig{figure=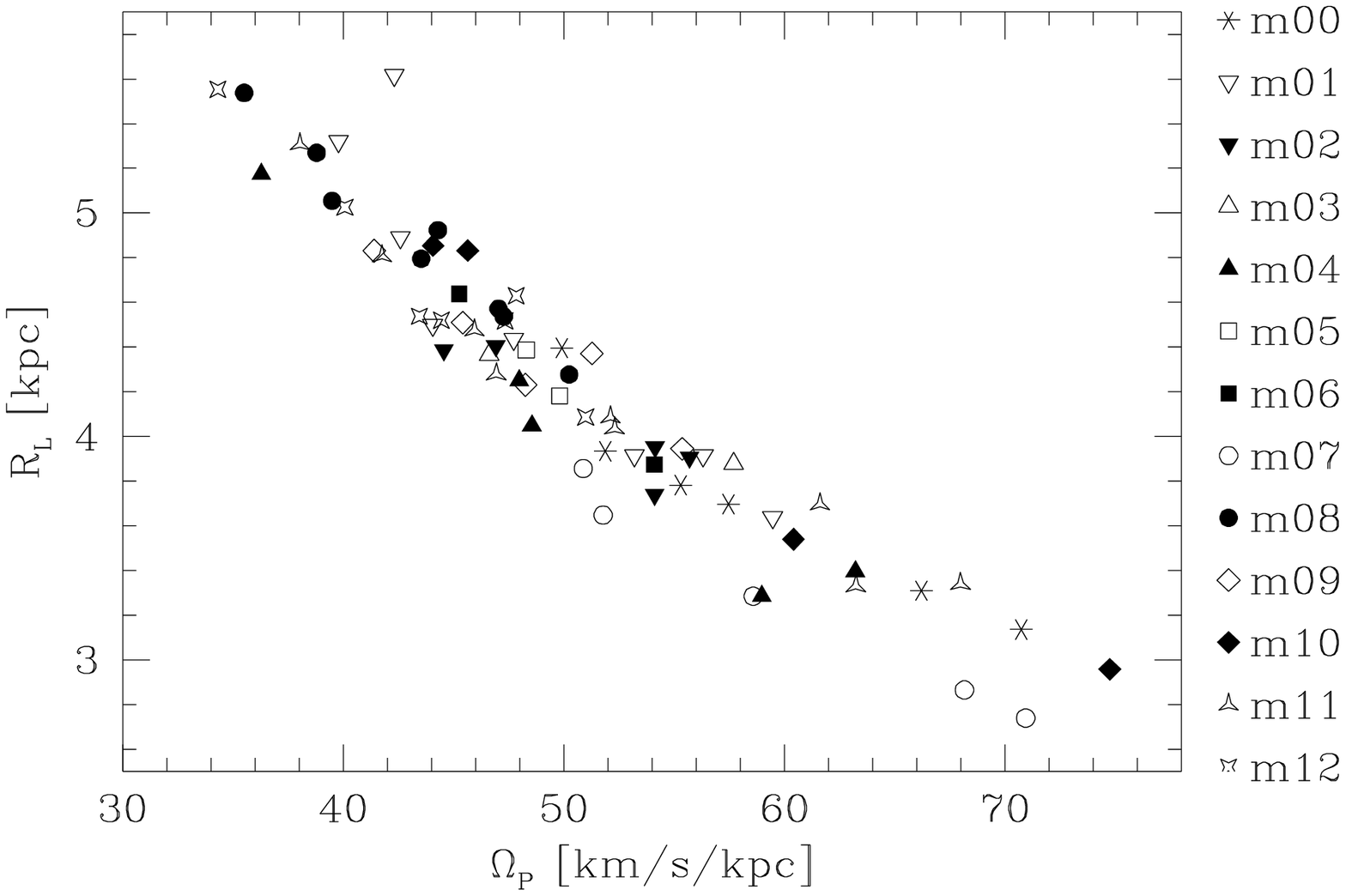,width=8.8cm}}
\caption[]{Corotation radius $R_{\rm L}$, derived as the distance of the
Lagrangian points $L_1$ and $L_2$ to the centre, versus the pattern speed of
the bar $\Omega_{\rm P}$ for the A~sample models. Each symbol refers to models
of the same simulation}
\label{rlo}
\end{figure}

%
An approximate bar pattern speed $\Omega_{\rm P}$ is derived for each model as
the least square slope of the bar position angle $\vartheta(t)$ relative to a
fixed axis, within a time interval of a few bar rotation and centred on the
current time. The related corotation radius $R_{\rm L}$ then follows from the
Lagrangian points $L_1$ and $L_2$, i.e. the saddle points of the effective
potential $\Phi_{\rm eff}=\Phi(R,z)-\frac{1}{2}R^2\Omega_{\rm P}^2$.
Figure~\ref{rlo} plots the sequence generated by our models in the
$\Omega_{\rm P}\!-\!R_{\rm L}$ plane. The B~sample values are quantified in
Table~\ref{loc} and the left frames in Fig.~\ref{qq} highlight the position
of the Lagrangian points. The B~sample models have
$\Omega_{\rm P}=50\pm 5$~km/s/kpc and $R_{\rm L}=4.3\pm 0.5$~kpc.
\par These results conflict with the dynamical properties
$\Omega_{\rm P}=63$~km/s/kpc and $R_{\rm L}=2.4\pm 0.5$~kpc reported by Binney
et al. (\cite{BG}) and are closer to $\Omega_{\rm P}=55$~km/s/kpc and
$R_{\rm L}=3.6$~kpc inferred by Weiner \& Sellwood (\cite{WS}) from SPH gas
flow modelling in a rigid potential. The former $\Omega_{\rm P}\!-\!R_{\rm L}$
pair moreover falls well below the relation displayed in Fig.~\ref{rlo}.
\par Binney et al. (\cite{BGS}) suggest that the peak emission in the COBE
near-IR maps at $b=0$ and near $l\!\sim\!-22\degr$, corresponding to
$R\approx 3$~kpc, is due to stars trapped by the Lagrangian point $L_4$/$L_5$.
We think that this peak could also arise from star formation in a gaseous
pseudo ring, possibly associated with the ``expanding 3 kpc arm'', seen
tangentially and located close to the inner 4:1 resonance, as the inner rings
in many external barred galaxies (Buta \cite{BU} and references therein).
According to this picture, corotation would lie significantly beyond 3 kpc.
\par Moreover, Zhao (\cite{Z}) has constructed a model for the Galactic bar
using Schwarzschild's technique and imposing $\Omega_{\rm P}=60$~km/s/kpc.
An $N$-body evolution of his model shows that this input pattern speed
decreases by about 20\% after one rotation period (see his Fig.~9), thus
becoming consistent with our results.
\par Binney et al. (\cite{BG}) have argued that prograde gas in a barred
potential, due to the pressure and viscous forces, settles onto the stable and
closed orbits of the $x_1$~family, switching progressively to ever lower
energy orbits as long as these develop no loops. Depending on the bar
potential, the gas may finally reach a last non-self intersecting,
``cusped'' $x_1$ orbit beyond which shocks transform most of the atomic gas
into molecular gas, and force the gas to plunge to the more viable orbits of
the $x_2$~family (if an inner Lindblad resonance exists).
Thus, atomic gas is expected to move along non-self intersecting $x_1$ orbits,
providing the key for a model comparison with Galactic HI observations.
In particular, the envelop defined by the traces of such orbits in the
longitude-velocity $l\!-\!V$~diagram should coincide with the observed HI
terminal velocities.
\par Orbits of the $x_1$ family have been computed in almost all our dynamical
models, using the instantaneous frozen potential. Figure~\ref{qq} shows the
comparison of several model $x_1$ traces versus HI contours in the
$l\!-\!V$~diagram. Some models reproduce fairly well the HI terminal velocity
envelop. The best cases certainly are m04t3000 and m06t4600, which share the
common property to arise from simulations with lower disc mass fraction in the
bar region: their disc to total mass ratio within $s<3$~kpc is less than 0.45,
where $s$ is the variable defined in Eq.~(\ref{s}). Further cases from the
B~sample are m12t2000, m08t3200, m09t1600, m02t2000 and m06t4800. All these
models, except m06t4800, occur shortly after the formation of the bar, when
redistribution of angular momentum has not completely disturbed the initial
exponential radial profile of the disc. In general, the presence of the bar
tends to steepen the azimuthally averaged inner radial profile of the disc,
increasing the central disc surface density.
\par Most of the models also have $x_1$ envelopes exceeding the observed
terminal velocities, indicating that there could be too much mass near the
centre. Such an excess could of course be reduced by increasing the
angle~$\varphi_{\circ}$, but then the peaks of maximum and minimum velocity
traced by the cusped $x_1$ orbit in the $l\!-\!V$~diagram would also be
shifted towards higher~$|l|$. Reducing the velocity scale in general render
the local disc kinematics less consistent with observations.
\par Hence, if the Galactic gas really moves on non self-intersecting closed
orbits, then the HI observations suggest that our dynamical models could have
too much mass in the disc near the centre.
\subsection{Mass-to-K luminosity ratio}
\label{mtl}
In addition to the position of the observer, the COBE-fits in~Sect.~\ref{adj}
also yield the $K$-band mass-to-light ratios $\Upsilon_K$ of the models,
without the contribution of the DH component.
\par Taking $\Upsilon_{K,\odot}=2.69\cdot 10^{-12}~M_{\odot}$W$^{-1}$Hz
(Wamsteker \cite{W}), the rescaled values for the B~sample models range from
$\Upsilon_K=0.53$ to 0.79 in Solar units, except for model m12t1600 (see
Table~\ref{loc}). The total NS+disc+DH mass of the B sample models within the
spheroid $s<3$~kpc is $(2.6\pm 0.15) \times 10^{10}~M_{\odot}$, with about 8\%
contribution from the DH component. This gives an average percentage by which
$\Upsilon_K$ is underestimated.
\par Modelling the gas dynamics of the barred galaxy M100 (NGC4321), which has
a Hubble type similar to the Milky Way, Knapen et al. (\cite{KB}) have
inferred $\Upsilon_K>0.7$ outside its nuclear ring. Moreover, for stellar
populations with near-IR emission dominated by late K and M giants, as for the
Galactic bulge (Arendt et al. \cite{AB}), similar mass-to-light ratios are
expected (Worthey \cite{WO}). Such lower limit for $\Upsilon_K$ rules out some
models, like those of simulation~m07 which all have $\Upsilon_K<0.4$. The
substantial microlensing optical depths towards the Galactic bulge also argue
for a large value of $\Upsilon_K$, as discussed below in Sect.~\ref{opt}.
\begin{figure}[t]
\centerline{\psfig{figure=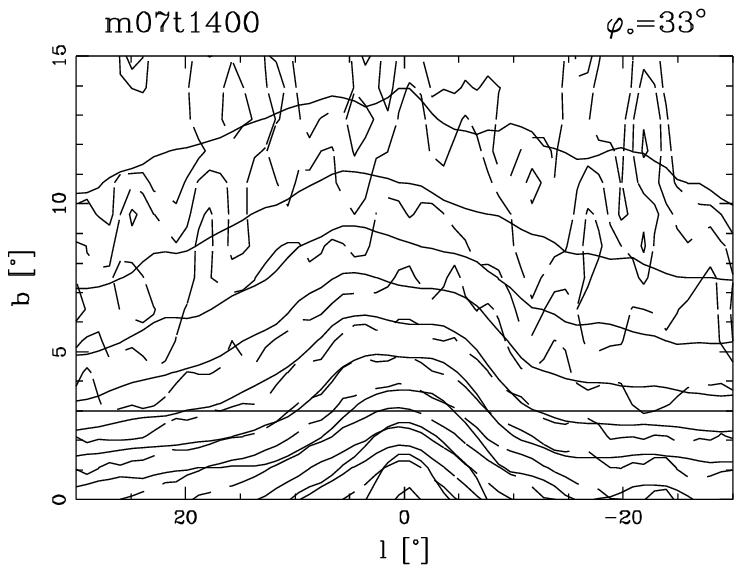,height=5.9cm}}
\caption[]{Confrontation of a model with small $h_z/R_{\circ}$ to the COBE
data. The caption is similar to Fig.~\ref{qq}}
\label{qqter}
\end{figure}
\par A noticeable difference between the COBE $K$-band map and the model maps
in Fig.~\ref{qq} is that the model contours are steeper in the low latitude
and $|l|\ga 15\degr$ region, i.e. where the disc becomes dominant. Even if
our correction for extinction fails at $b\la 3\degr$, the near-IR contours
in this region still remain very flat after more elaborated dust subtraction
of the COBE data (Spergel et al. \cite{SM}) and hence the difference is
probably real. Several reasons may lead to this departure.
\par First, the Galactic disc scale height may be lower by a factor $\sim 2$
in the inner Galaxy than in the Solar neighbourhood, as deduced by Kent et al.
(\cite{KDF}) from the Spacelab IR Telescope data and by Binney et al.
(\cite{BGS}) from the deprojection of the COBE $L$-band map. At constant
surface density, discs with smaller scale height are more concentrated towards
the plane and should therefore contribute more to the low latitude near-IR
emission. Our models, started with radially constant disc thickness, do not
present such large disc scale height gradient outside the bar region. But
models from simulation~m07 (see Fig.~\ref{qqter}), with the thinnest disc,
indeed have flatter $l\!-\!b$ contours than models from m08. The variable
scale height alternative however is not supported by observations in external
late-type spiral galaxies (de Grijs \& van der Kruit \cite{GK} and reference
therein).
\par Second, discs with more foreground mass between the bulge and the observer
will also enhance the low latitude integrated light. This is the case for the
m06~models (Fig.~\ref{qq}). At fixed total disc mass, because of the higher
initial disc scale length, these models have higher disc surface density in
the outer region, including the first few kpc below $R_{\circ}$. Increasing
simply the disc mass is less efficient because the inner disc, mixing with the
NS component during the bar instability, will also contribute more to the bulge
emission.
\begin{figure*}[t]
\centerline{\psfig{figure=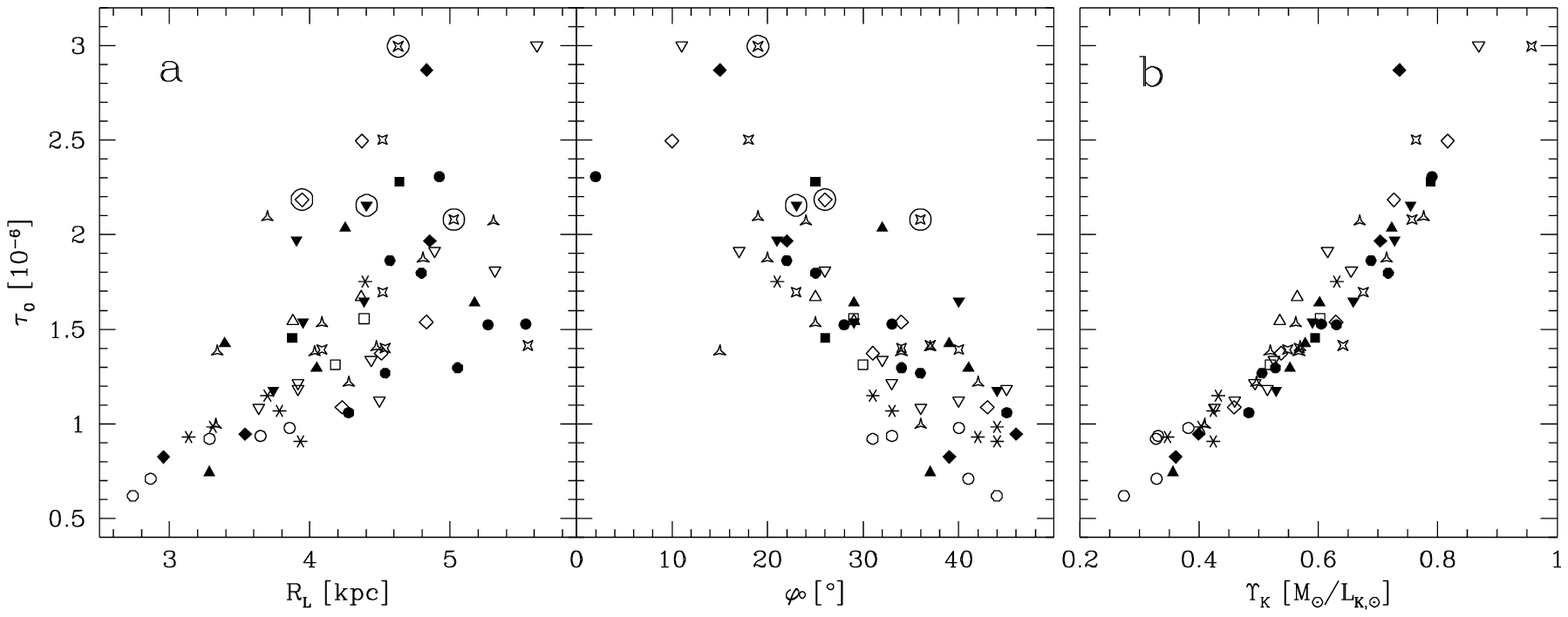,width=18cm}}
\caption[]{{\bf a} Dependence of the microlensing optical depths $\tau_0$
(without DH contribution) towards Baade's Window on the corotation radius
$R_{\rm L}$, which is a reasonable indicator of the radial extension of the
bar, and on the bar inclination angle $\varphi_{\circ}$, for the A~sample
models. The encircled points indicate models with prominent spiral structures.
{\bf b} Relation between $\tau_0$ and the mass-to-light ratio $\Upsilon_K$ for
the same models. The symbols are as in Fig.~\ref{rlo}}
\label{prt}
\end{figure*}
\par Finally, the assumption $\Upsilon_K\!={\rm const.}$ may not hold close to
the Galactic plane because of near-IR emission from interstellar matter, or
recent star formation. In the latter case, red supergiants, too young for
having diffused as far above the plane as the old disc population, could
significantly contribute to the $K$-emission and reduce $\Upsilon_K$ at low
Galactic latitude. However our COBE-adjustments should not be affected by such
a $\Upsilon_K$ gradient since the region $|b|<3\degr$ was excluded.
Alternatively, the IR mass-to-light ratio could also be lower in the
Galactic disc than in the bulge, as found by Verdes-Montenegro et al.
(\cite{VM}) for NGC7217 in the $I$-band.
\par In the next section, we will argue for the massive disc possibility
and against a variable disc scale height.
\subsection{Bulge microlensing}
\label{opt}
The microlensing optical depths $\tau$ towards the galactic bulge provide a
serious direct probe of the mass distribution inside the Solar circle.
\par The first results obtained by the OGLE and MACHO experiments favour
surprisingly large optical depths: the reported values are {\it at least}
$(3.3\pm 1.2)\times 10^{-6}$ for 9 bulge stars essentially located in
Baade's Window (Udalski et~al. \cite{U}), and $(3.9_{-1.2}^{+1.8})\times
10^{-6}$ for 13 clump giants with mean Galactic coordinates $l=2\fdg 55$
and $b=-3\fdg 64$ (Alcock et al. \cite{AA}), whereas axisymmetric models
predict values less than $10^{-6}$ (Evans \cite{E} and references therein).
The contribution of bulge lenses may however significantly increase the model
predictions if the bulge is an elongated bar seen nearly end-on (Evans
\cite{E}, Kiraga \& Paczynsky \cite{KP}, Zhao et al. \cite{ZR}, Zhao \& Mao
\cite{ZM}).
\par We have computed the model microlensing optical depths towards Baade's
Window, averaged over all sources along the line-of-sight, from the following
Monte Carlo version of Eq.~(5) given by Kiraga \& Paczynski (\cite{KP}):
\begin{equation}
\tau_{\beta\!_s}\!=\!\frac{4\pi G}{\Delta\Omega c^2}\!\left[\sum_{D_{\rm s}}
\!\sum_{D_{\rm d}<D_{\rm s}}\!\!\left(\frac{m_{\rm d}}{D_{\rm d}}\!-
\!\frac{m_{\rm d}}{D_{\rm s}}\right)D_{\rm s}^{2\beta\!_s}\!\right]\!
\!\left[\sum_{D_{\rm s}}D_{\rm s}^{2\beta\!_s}\!\right]^{-1}
\label{tau}
\end{equation}
where the outer sums involve all source particles within a solid angle $\Delta
\Omega$ of the selected direction and the inner sum all lens particles in the
same solid angle between the specified source and the observer, the symbols
$D_{\rm s}$ and $D_{\rm d}$ referring resp. to the source and lens
distances relative to the observer, and $m_{\rm d}$ to the mass of the lens
particles. $G$~is the gravitational constant, $c$ the speed of light and
$\beta_s$ the parameter introduced by Kiraga \& Paczynski (\cite{KP}) to
describe the detection probability of the sources: $\beta_s=0$ if the sources
are detectable whatever their distance, like clump giant stars, and
$\beta_s\approx -1$ for main sequence stars (Bissantz et al. \cite{BE} and
references therein).
\par Optical depths are calculated with and without including DH lenses.
Table~\ref{out} gives some results for $\beta_s=0$. The optical depths for
other values of $\beta_s$ are almost proportional to $\tau_0$: in particular,
we find $\tau_{-1}\approx \frac{2}{3} \tau_{0}$, both with or without DH
lenses. The DH contributes roughly 9\% to the total optical depth. In the mean
direction of the \mbox{Alcock} et al. (\cite{AA}) fields, $\tau_{0}$ is on the
average 15\% higher than in Baade's Window, in agreement with the above
reported observations.
\begin{figure}[t]
\centerline{\psfig{figure=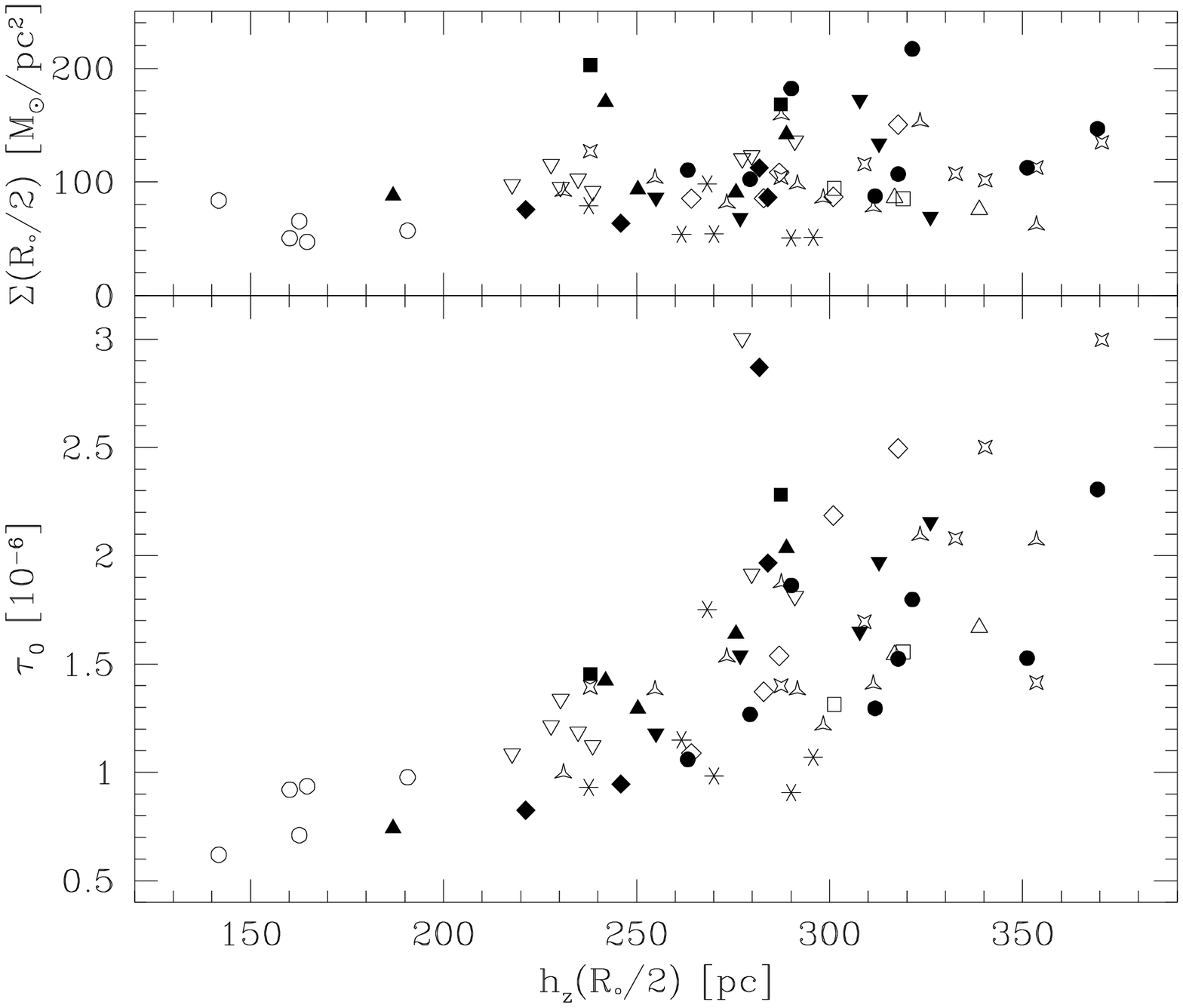,width=8.8cm}}
\caption[]{Microlensing optical depths $\tau_{0}$ (without DH) towards Baade's
Window versus disc scale height half way between the observer and the Galactic
centre for the A~sample models. The upper frame shows that the correlation
is not induced by a variable disc surface density. The symbols are as in
Fig.~\ref{rlo}}
\label{hz}
\end{figure}
\par Figure~\ref{prt}a shows the optical depths of the A~sample models towards
Baade's Window as a function of the size of the bar and its inclination angle
relative to the observer. The optical depth strongly depends on
$\varphi_{\circ}$, increasing from $\tau_0\sim 10^{-6}$ for $\varphi_{\circ}=
40\degr$ to $\tau_0\sim 2.5\times 10^{-6}$ for $\varphi_{\circ}=15\degr$, and
is roughly proportional to $R_{\rm L}$. The dispersion in each plot partly
reflects the dependence of the optical depth on both $\varphi_{\circ}$ and
$R_{\rm L}$: at fixed $R_{\rm L}$, the larger values of $\tau_0$ come from
models with more end-on bars.
\par Clearly, the optical depths depends on the mass scaling and thus on the
adopted mass-to-light ratio $\Upsilon_K$, as depicted in Fig.~\ref{prt}b.
Within the various investigated models, this relation does not seem to depend
much on a third parameter. According to it, the $1\sigma$ lower limit of the
observed optical depths, $\tau\ga 2.1\times 10^{-6}$, implies $\Upsilon_K>0.7$,
consistent with the previous discussion on this parameter (Sect.~\ref{mtl}).
\par If more event statistics confirm the large observed optical depths, then
the models in Fig.~\ref{prt}a with our best COBE-estimate $28\degr \pm 7\degr$
for the bar inclination angle are inconsistent with the microlensing
constraints. This could indicate that our models have insufficient mass along
the line-of-sight in Baade's Window. To enhance that mass without increasing
the $x_1$-terminal velocities, an obvious possibility is to increase the disc
mass outside the bar region at the expense of the DH, i.e. approaching a
maximum disc solution, as suggested by Alcock at al. (\cite{AA}) and by the
recent Galactic structure review of Sackett (\cite{S}). If dark matter in the
outer Galaxy is in molecular form (Pfenniger et al. \cite{PC}) and thus
concentrated in the plane, its contribution to the squared inner circular
velocity can become {\it negative}, even allowing for an over maximum disc
solution. To ensure a low central surface density, such a heavy disc would
then have to be partially hollow relative to its exponential extrapolation
inside the bulge.
\par Microlensing optical depths towards Baade's Window also depend on the disc
scale height. At fixed surface density and distance $z_1$ above the plane, one
can show that the (exponential) scale height which maximises the volume density
is $h_z=z_1$ if the vertical mass distribution is exponential. Furthermore,
the lenses which most contribute to the optical depth are those lying half way
between the source and the observer and, for reasonable disc parameters, the
mass density along Baade's Window is about constant (e.g. for $h_R=3.5$~kpc and
$h_z(R)={\rm const.}=240$~pc). Thus for bulge sources, the maximum disc
contribution in this window should come from stars at $R=4\!-\!5$~kpc, which
are about 270~pc away from the plane. Our models indeed produce larger values
of $\tau_0$ on the average when $h_z(R_{\circ}/2)\approx 300$~pc, as shown in
Fig.~\ref{hz}. In particular, models with $h_z\la 250$~pc in the inner part
all have $\tau_0<1.5\times 10^{-6}$, hard to conceal with observations. From
this we infer that the mass distribution of the Galactic disc probably has a
constant scale height between $\sim R_{\circ}/2$ and $R_{\circ}$, contrary to
the constant mass-to-light ratio interpretations of near-IR surface photometry
data (Kent et al. \cite{KDF}; Binney et al. \cite{BGS}).
\begin{figure}[t]
\centerline{\psfig{figure=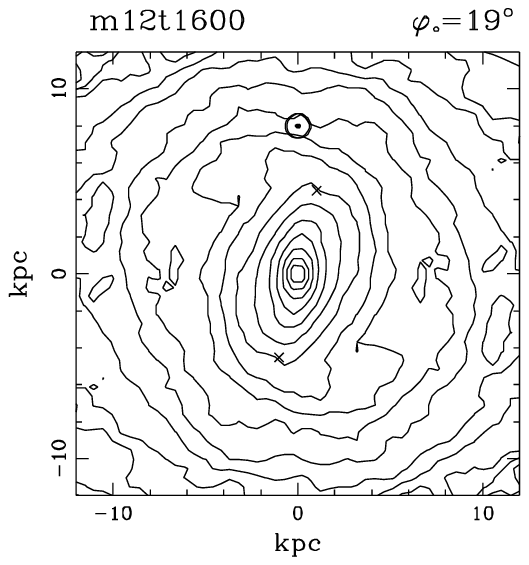,height=5.9cm}}
\caption[]{Face-on view of a model with optical depth towards Baade's Window
over $3\times 10^{-6}$. The caption is similar to Fig.~\ref{qq} and the model
has been symmetriesed}
\label{qqbis}
\end{figure}
\par The effect of asymmetries like spiral arms may also affect the optical
depths towards the bulge, as illustrated in Fig.~\ref{prt}a. Model m12t1600
has one of the largest $\tau_0$ towards Baade's Window and indeed exhibits a
strong spiral structure (see Fig.~\ref{qqbis}). Another example, with larger
$\varphi_{\circ}$ and smaller $R_{\rm L}$, is m09t1600 (see Fig.~\ref{qq}).
The orientation of the Galactic bar is such that the line-of-sight through
Baade's Window crosses the spiral arm starting at the near end of the bar
almost tangentially and at a distance where microlensing should be very
efficient. The gain in optical depth is of order $0.5 \times 10^{-6}$.
\renewcommand{\thefigure}{\arabic{figure}a-d}
\begin{figure*}[t]
\centerline{\psfig{figure=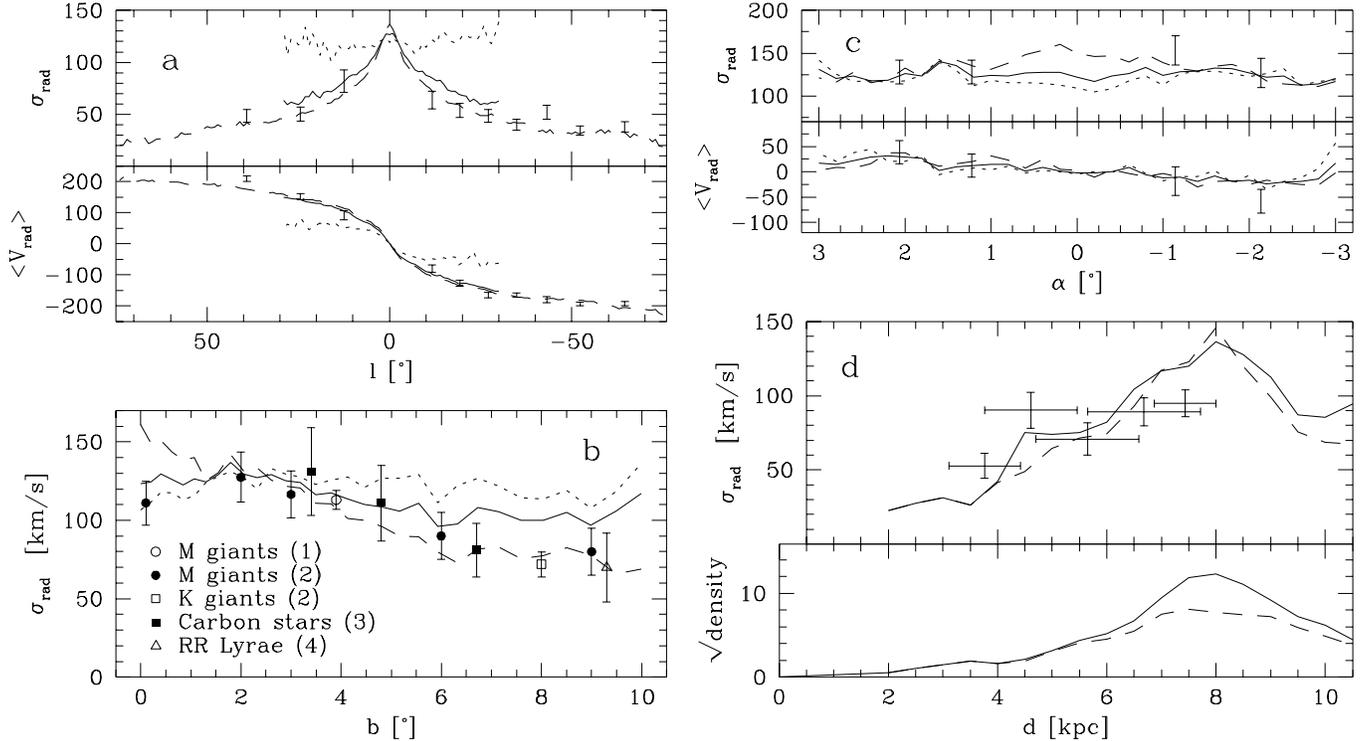,width=18cm}}
\caption[]{Galactocentric radial kinematical properties of the model m08t3200.
In all plots, the  dashed, dotted and full lines stand resp. for disc,
NS and disc+NS particles. {\bf a}~Radial velocity dispersion and mean velocity
within $|b|<5\degr$ based on particles interior to the Solar circle. The points
with error bars are derived from the te Lintel Hekkert et al. (\cite{TC})
catalogue of double peaked OH/IR stars. {\bf b}~Projected radial velocity
dispersion along the bar minor axis ($l=0$). The data are from: (1)~Sharples
et al. \cite{SW}; (2)~Terndrup published by Rich \cite{R}; (3)~Tyson \& Rich
\cite{TR}; (4)~Rodgers \cite{RO}. {\bf c}~Projected radial velocity moments
along an axis through the Galactic centre and inclined by $55\degr$
relative to the minor axis, and the corresponding Blum et al. (\cite{BC})
M giant observations. The angle $\alpha$ is measured with respect to $l=b=0$
and has the same sign as $l$. {\bf d}~Radial velocity~dispersion in Baade's
Window as a function of the distance $d$ from the observer, with the K giant
data of Lewis \& Freeman (\cite{LF}). The lower plot also represents the
square root of the relative mass distribution per solid angle along the
line-of-sight (arbitrary~units)}
\label{IK}
\end{figure*}
\renewcommand{\thefigure}{\arabic{figure}}
\subsection{Inner kinematics}
%
It is beyond the scope of this paper to study the detailed kinematics of all
the dynamical models. Instead, we present here the case of a single model,
m08t3200, which is in fair agreement with almost all explored constraints.
In Figs.~\ref{IK}, several kinematical predictions of this model are compared
to stellar observations.
\par The agreement between the longitudinal kinematics of the model disc and
that of the OH/IR stars (Fig.~16a) is very convincing. The model
particles outside the Solar circle were not taken into account as the data
contain only very few OH/IR stars at $|l|>90\degr$. No velocity moment has
been derived from the data within $|l|<10\degr$ because in this region the
radial velocity dispersion depends non-negligibly on the galactic latitude
and the te~Lintel~Hekkert et al. (\cite{TC}) OH/IR stars severely suffer from
undersampling at $b\la 2\degr$.
\par The minor axis velocity dispersion of the model (Fig.~16b) is
different for the NS and disc components. The observed kinematics are best
traced by the NS component near the centre and by the disc component at
large Galactic latitude. In the Blum et al. (\cite{BC}) fields, the
kinematics of these components are more similar (Fig.~16c).
\par The velocity dispersion in Baade's Window as a function of the distance
from the observer is close to that observed for K giants (Fig.~16d),
except near the tangent point. However, the projected velocity dispersion of
these stars is obviously less than the 113 km/s of the M giants which was used
to scale the model. Furthermore, the distance distribution of the foreground
model particles is roughly proportional to $d\,^2$, whereas the distribution
of the K giants seems nearly linear with $d$ (Sadler et al. \cite{SR}),
suggesting that the latter may be biased towards the near stars. This
difference between model and observations is hardly due to a variable Galactic
disc scale height. One can indeed show that, within $3\!-\!4$~kpc from the
observer, the simulated line-of-sight mass distribution towards Baade's Window
in a realistic analytical double exponential disc is very similar for a
radially constant or a linearly increasing $h_z$ if the face-on surface
density profile is kept the same.
\begin{table}[t]
\centering
\caption[]{Comparison of model m08t3200 versus K giant projected radial
kinematics in off axis bulge fields. Boldfaced, italic and roman velocities
refer resp. to disc, NS and disc+NS particles. The mean velocities are
Galactocentric. The references for the observations are Harding \cite{HP},
Minniti \cite{M1}, Minniti \cite{M2} and Minniti et al. \cite{MW}}
\begin{tabular}{rrcrrrr} \hline
&&& \multicolumn{2}{c}{Observations} & \multicolumn{2}{c}{Model} \\
$l [\degr]$ & $b [\degr]$ & [Fe/H] & \multicolumn{1}{c}{$\sigma_r$} &
\multicolumn{1}{c}{$\overline{v_r}$} & $\sigma_r$ & $\overline{v_r}$ \\ \hline
 -10.0 & -10.0 & $>-1$  & $ 67\pm \;\;6$ & $-82\pm \;\;8$ &  {\bf 72} &
                                                                  {\bf -81} \\
       &       & $<-1$  & $107\pm \;\;6$ & $-37\pm \;\;8$ & {\it 120} &
                                                                  {\it -41} \\
       &       &  all   & \multicolumn{1}{c}{--} & \multicolumn{1}{c}{--} &
                                                           110  &      -71  \\
   9.9 &  -7.6 & $>-1$  & $ 70\pm \;\;7$ & $ 56\pm 10$    &  {\bf 70} &
                                                                   {\bf 92} \\
       &       & $<-1$  & $ 91\pm 13$    & $-18\pm 18$    & {\it 117} &
                                                                   {\it 40} \\
       &       &  all   & \multicolumn{1}{c}{--} & \multicolumn{1}{c}{--} &
                                                            99  &       66  \\
   8.0 &   7.0 & $>-1$  & $ 72\pm \;\;4$ & $ 66\pm \;\;5$ &  {\bf 77} &
                                                                   {\bf 68} \\
       &       & $<-1$  & $109\pm 10$    & $ -7\pm 14$    & {\it 121} &
                                                                   {\it 36} \\
       &       &  all   & \multicolumn{1}{c}{--} & \multicolumn{1}{c}{--} &
                                                           100  &       54  \\
  12.0 &   3.0 &   --   & \multicolumn{1}{c}{--} & \multicolumn{1}{c}{--} &
                                                       {\bf 75} & {\bf 111} \\
       &       &   --   & \multicolumn{1}{c}{--} & \multicolumn{1}{c}{--} &
                                                      {\it 124} &  {\it 45} \\
       &       &  all   & $ 68\pm \;\;6$ & $ 77\pm \;\;9$ &        93 &
       96  \\
                                                                        \hline
\end{tabular}
\label{VW}
\end{table}
\par Table~\ref{VW} reviews some off-axis K giant observations and give the
corresponding model predictions. The velocity moments of the model disc
component resemble those of the giants with [Fe/H]$>-1$.
The $l\!-\!b$ proper motion dispersions of the model in Baade's Window
are $(\sigma_{\mu_l},\sigma_{\mu_b})=(3.15,2.38)$~m$\arcsec$/yr for the disc,
$(2.96,2.81)$~m$\arcsec$/yr for the NS and $(3.08,2.57)$~m$\arcsec$/yr for
both visible components together, whereas the observed values for K giants are
$(3.2\pm 0.1,2.8\pm 0.1)$~m$\arcsec$/yr
(Spaenhauer et al. \cite{SJ}).
%
\section{Conclusion}
We have built many self-consistent 3D dynamical barred models of the Milky Way
extending beyond $R=R_{\circ}$ by $N$-body integration of various bar
unstable axisymmetric models. The models, extracted from the simulations at
a frequency of 200 Myr, include 3 components: a nucleus-spheroid standing for
the Galactic inner bulge and stellar halo, a disc mainly representing the
Galactic old disc and a non-dissipative dark halo. The comparison of the
models with observational constraints leads to the following considerations.
\par 1) The spatial location of the observer in each model is constrained by
the COBE/DIRBE $K$-band map corrected for extinction by dust, assuming a
constant mass-to-light ratio $\Upsilon_K$ for the luminous mass components.
The results for the best matching models, with mean quadratic residuals
between model and data fluxes down to $0.3$\%, suggest that the angle between
the $l=b=0$ line and the major axis of the Galactic bar is $28\degr\pm 7\degr$.
\par 2) Scaling the models such as the distance of the observer to the centre
is $R_{\circ}=8$~kpc and the projected radial velocity dispersion towards
Baade's Window 113 km/s, as observed for M giants, absolute model properties
are derived. A dozen of models reproduce fairly well both the COBE-data and
observations in the Solar Neighbourhood, although with a rather low radial
versus azimuthal velocity dispersion anisotropy of the spheroid components.
\par 3) The bars in these models have a face-on axis ratio $b/a=0.5\pm 0.1$
and a pattern speed $\Omega_{\rm P}=50\pm 5$~km/s/kpc, placing the corotation
at $4.3\pm 0.5$~kpc. Models with a disc mass fraction below 0.45 within
3~kpc from the centre produce envelops of non self-intersecting $x_1$-orbits
in the $l\!-\!V$~diagram which better agree with the observed HI terminal
velocities.
\par 4) The microlensing optical depths of the models towards the bulge
strongly depends on the bar inclination angle $\varphi_{\circ}$, increasing
from $\tau_0\sim 10^{-6}$ for $\varphi_{\circ}=40\degr$ to $\tau_0\sim
2.5\times 10^{-6}$ for $\varphi=15\degr$ towards Baade's Window, whereas
observations rather support values over $3\times 10^{-6}$. We find that a
spiral arm starting at the near end of the bar can increase the optical depths
by $0.5\times 10^{-6}$ and thus reduce the gap between model and observed
values.
\par 5) All models with a disc scale height $h_z\leq 250$~pc half way between
the observer and the Galactic centre have $\tau_0<1.5\times 10^{-6}$, arguing
against an inwards decreasing disc scale height. This result is also in
agreement with the constant $h_z$ observed in late-type spirals.
\par 6) The models predict a mass-to-$K$ luminosity ratio $\Upsilon_K=
0.6\!-\!0.8$ in Solar units. Values near the upper limit are consistent with
the mass-to-light ratio estimated for M100 (NGC4321), a galaxy similar to the
Milky Way, and favour microlensing optical depths closer to the observed
values. There is indeed an obvious but tight correlation between $\tau_0$ and
$\Upsilon_K$, according to which $\tau_0\ga 2\times 10^{-6}$ implies
$\Upsilon_K>0.7$.
\par 7) Most models predict too low surface brightness relative to the
COBE-data in the region $|l|\la 15\degr$ dominated by the disc. Beside an
improbable lower disc scale height in the inner Galaxy, or a variable
mass-to-light ratio, this discrepancy could indicate that the Galactic disc
outside the bar region is more massive than assumed in the models, favouring
large microlensing optical depths and arguing for a maximum disc Milky Way.
\par 8) The disc radial kinematics of the models resembles the observed
kinematics of K giants with ${\rm [Fe/H]}>-1$ in the outer bulge.
\par As reasonable models regarding most observational constraints, we would
recommend the models m08t3200 and m04t3000.
\acknowledgements{I would like to thank L.~Martinet and D.~Pfenniger for many
enlightening discussions and for critical reading of the paper, as well as
J.~Sellwood for refering this paper and D.~Pfenniger for providing its
efficient $N$-body code. This work has been partially supported by the Swiss
National Science Foundation.}
\normalsize
\appendix
\section*{Appendix A: a non-Gaussian bounded 3D distribution}
\setcounter{section}{1}
\setcounter{figure}{0}
\renewcommand{\thefigure}{\thesection\arabic{figure}}
A convenient 3D distribution with non-zero probability over a finite volume,
i.e. avoiding the tails of the multi-normal distribution, is given by:
\begin{equation}
\begin{array}{l}
B_3(\xi,\eta,\zeta) = \left[2^{\kappa+\lambda-1}B(\kappa,\lambda)
  B(\frac{1}{2},\mu)B(\frac{1}{2},\omega)\right]^{-1}\cdot \\ \\
\left\{\begin{array}{cl}
       \multicolumn{2}{l}{\!(1\!+\!\xi)^{\kappa-2}(1\!-\!\xi)^{\lambda-2}
       (1\!-\!\frac{\eta^2}{1\!-\xi^2})^{\mu-\frac{3}{2}} \vspace{.2cm}
       (1\!-\!\frac{\zeta^2}{1\!-\xi^2\!-\eta^2})^{\omega-1}} \\
       & \hbox{if} \;\; \xi^2+\eta^2+\zeta^2 \leq 1 \\ \\
       \hspace{1cm} 0 \hspace{3.2cm}& \hbox{otherwise},
       \end{array} \right.
\end{array}
\label{B3}
\end{equation}
where $B$ is the Beta function, and $\kappa$, $\lambda$, $\mu$ and $\omega$
are four parameters. In the reduced variables $\xi$, $\eta$ and $\zeta$, this
distribution is bounded by a sphere of radius 1 on which, as long as $\kappa,
\lambda>2$, $\mu>3/2$ and $\omega>1$, it continuously vanishes. If $\kappa=
\lambda=2$, $\mu=3/2$ and $\omega=1$, the distribution is homogeneous inside
the boundary sphere and smaller values of these parameters produce
singularities on it.
\par
\begin{figure}
\centerline{\psfig{figure=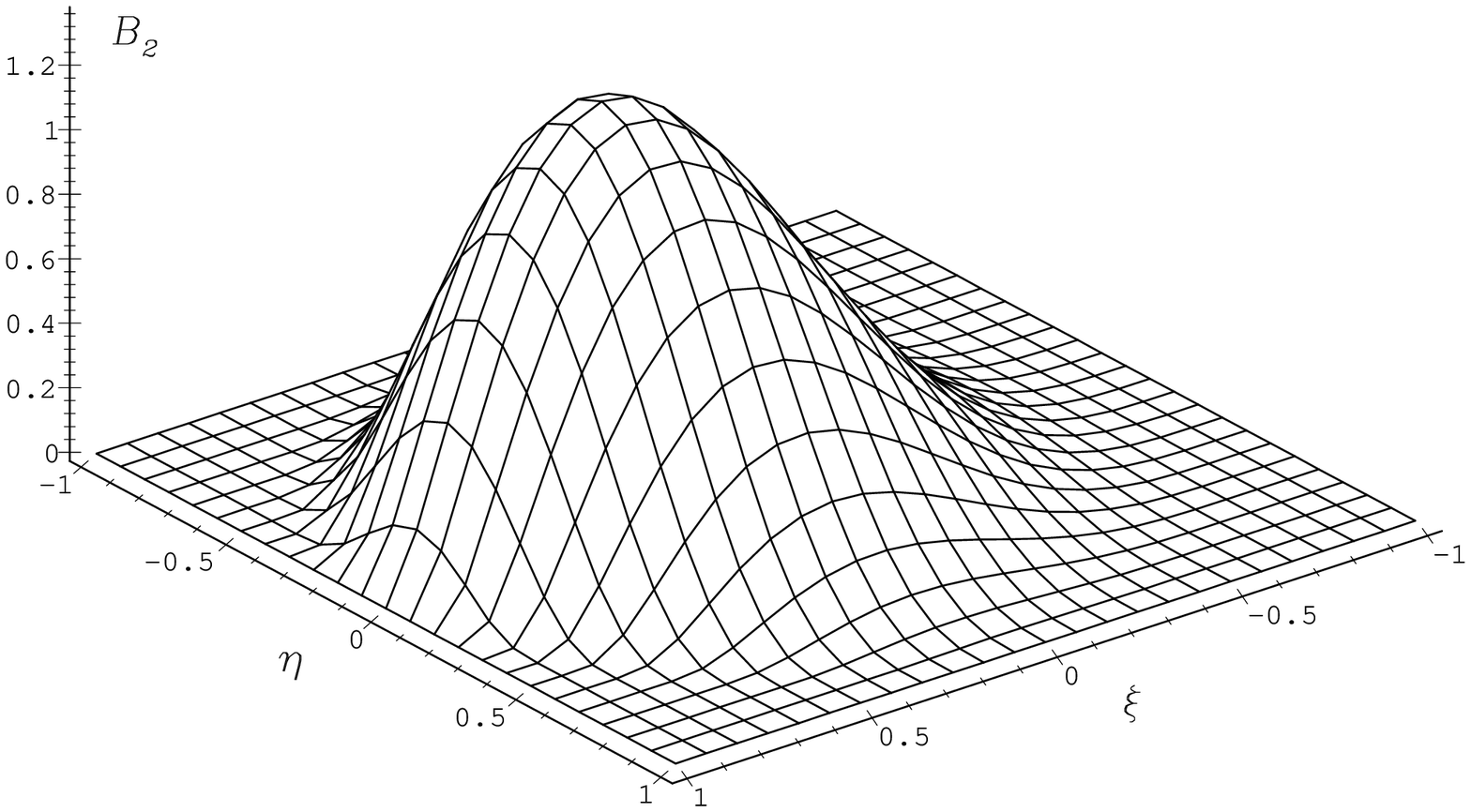,width=8.8cm}}
\caption[]{Distribution of the variables $\xi$ and $\eta$ resulting from
Eq.~(\ref{B2}), with $\kappa=5$, $\lambda=3$ and $\mu=4$}
\label{beta}
\end{figure}

 The first and second moments are:
\begin{equation}
\overline{\eta}=\overline{\zeta}=\overline{(\xi-\overline{\xi})\eta}
               =\overline{(\xi-\overline{\xi})\zeta}=\overline{\eta\zeta}=0,
\label{zero}
\end{equation}
\begin{eqnarray}
\overline{\xi} & = & \frac{\kappa-\lambda}{\kappa+\lambda}, \label{xm} \\
\sigma_{\xi\xi}^2 & \equiv & \overline{(\xi-\overline{\xi})^2}=
     4\frac{\kappa\lambda}{(\kappa+\lambda)^2(\kappa+\lambda+1)}, \label{sx} \\
\sigma_{\eta\eta}^2 & \equiv & \overline{\eta^2}=\frac{4}{2\mu+1}
      \frac{\kappa\lambda}{(\kappa+\lambda)(\kappa+\lambda+1)}, \label{sy} \\
\sigma_{\zeta\zeta}^2 & \equiv & \overline{\zeta^2}=\sigma_{\eta\eta}^2
      \cdot\frac{2\mu}{2\omega+1}. \label{sz}
\end{eqnarray}
Depending on $\kappa$ and $\lambda$, the distribution is skewed in $\xi$, the
maximum of probability lying at:
\begin{equation}
\xi_{\rm max}=\frac{\kappa-\lambda}{\kappa+\lambda-4}.
\label{xmax}
\end{equation}
Figure~\ref{beta} displays an example of the 2D distribution $B_2$ obtained
after integrating $B_3$ over $\zeta$:
\begin{eqnarray}
B_2(\xi,\eta) & \equiv & \int B_3(\xi,\eta,\zeta)d\zeta \nonumber \\
              & \propto & \! (1\!+\!\xi)^{\kappa-3/2}(1\!-\!\xi)^{\lambda-3/2}
     \!\left(1\!-\!\frac{\eta^2}{1\!-\!\xi^2}\right)^{\! \mu-1}\hspace{-.5cm}.
\label{B2}
\end{eqnarray}
\par The inversion of Eqs.~(\ref{xm})-(\ref{sz}) provides the parameters of
the \mbox{$B_3$-distribution} as a function of the aimed moments:
\begin{eqnarray}
\kappa  & = & \frac{1}{2} \frac{(1+\overline{\xi})(1-\overline{\xi}^2
              -\sigma_{\xi \xi}^2)}{\sigma_{\xi \xi}^2}, \label{kappa} \\
\lambda & = & \frac{1}{2} \frac{(1-\overline{\xi})(1-\overline{\xi}^2
              -\sigma_{\xi \xi}^2)}{\sigma_{\xi \xi}^2}, \label{lambda} \\
\mu     & = & \frac{1}{2} \frac{1-\overline{\xi}^2-\sigma_{\xi \xi}^2
              -\sigma_{\eta \eta}^2}{\sigma_{\eta \eta}^2}, \label{mu} \\
\omega  & = & \frac{1}{2} \frac{1-\overline{\xi}^2-\sigma_{\xi \xi}^2
              -\sigma_{\eta \eta}^2-\sigma_{\zeta \zeta}^2}
              {\sigma_{\zeta \zeta}^2}. \label{omega}
\end{eqnarray}
To generate random numbers distributed according to $B_3$, one may use the
property that:
\begin{eqnarray}
t_{\xi}   & \equiv & \frac{1}{2}(\xi+1), \label{tx} \\
t_{\eta}  & \equiv & \frac{1}{2}(\frac{\eta}{\sqrt{1-\xi^2}}+1), \label{ty} \\
t_{\zeta} & \equiv & \frac{1}{2}(\frac{\zeta}{\sqrt{1-\xi^2-\eta^2}}+1)
\label{tz}
\end{eqnarray}
follow Beta distributions with parameters $\kappa,\lambda$ for $t_{\xi}$,
$\gamma,\gamma$ for $t_{\eta}$, and $\omega,\omega$ for $t_{\zeta}$.
\section*{Appendix B: the ${\cal R}^2$ residual}
\setcounter{section}{2}
\setcounter{equation}{0}
If the model fluxes are decomposed into $F_i=F_{i,1}+F_{i,2}$, where $F_{i,1}$
is the exact part and $F_{i,2}$ is the statistical error due to the finite
number of particles, the $\chi^2$ in Eq.~(\ref{chi2}) expands into:
\begin{eqnarray}
\chi^2 & = & \sum_{i=1}^{N_{\rm pix}}\frac{(F_{i,1}-F_i^{\circ})^2}
                                   {\sigma_i^2} \nonumber \\
       &   & +\sum_{i=1}^{N_{\rm pix}}\frac{F_{i,2}^2}{\sigma_i^2}
     +2\sum_{i=1}^{N_{\rm pix}}\frac{F_{i,2}(F_{i,1}-F_i^{\circ})}{\sigma_i^2}.
\label{dchi2}
\end{eqnarray}
The expected value of the second term in the right hand side, when resulting
from a fit, is the number of degree of freedom $\nu$ (if the errors were
Gaussian, then this term would follow the standard $\chi^2$~statistics),
whereas the last term is about zero since by definition the $F_{i,2}$'s must
vanish on the average. With these simplifications, the best estimate of the
function:
\begin{equation}
{\cal R}^2\equiv\left[\sum_{i=1}^{N_{\rm pix}}
\frac{1}{\sigma_i^2/{F_i^{\circ}}^2}\right]^{-1}\sum_{i=1}^{N_{\rm pix}}
\frac{[(F_{i,1}-F_i^{\circ})/F_i^{\circ}]^2}{\sigma_i^2/{F_i^{\circ}}^2},
\end{equation}
i.e. the quadratic relative residuals between the $F_{i,1}$'s and
the $F_i^{\circ}$'s averaged over pixels and weighted by the inverse of the
relative variance, indeed reduces to Eq.~(\ref{calR}).
\section*{Appendix C: the variance of the model fluxes}
\setcounter{section}{3}
\setcounter{equation}{0}
The model flux in a given pixel is of the form:
\begin{equation}
F(\vec{D}) = \sum_{k=1}^{N}f(D_k),
\end{equation}
where $N$ is the total number of particles in the pixel, the $D_k$'s are the
distances of the particles relative to the observer and $\vec{D}\equiv
(D_1,\ldots,D_N)$. Noting resp. $P(N)$ and $p(D)$ the distributions of
$N$ and of the distances, the two first moments of $F$ are:
\begin{eqnarray}
\overline{F^n} & \!\equiv & \sum_{N=0}^{\infty}\! P(N) \!
                     \cdot \! \int_0^{\infty}\! dD_1 p(D_1) \! \ldots \!
                     \int_0^{\infty}\! dD_N p(D_N) F(\vec{D})^n \nonumber \\
& = &\left\{\begin{array}{ll}
 \overline{N}\cdot \overline{f} \vspace{.3cm} & \hspace{1cm} n=1 \\
 \overline{N}\,\overline{f^2}+(\overline{N}\,\overline{f})^2 & \hspace{1cm}n=2,
            \end{array}\right.
\label{mom}
\end{eqnarray}
where
\begin{eqnarray}
\overline{N} & = & \sum_{N=0}^{\infty} N P(N), \\
\overline{f^n} & = & \int_{0}^{\infty} p(D) f(D)^n dD \hspace{1cm} n=1,2.
\end{eqnarray}
The result for $n=2$ in Eq.~(\ref{mom}) requires the assumption that the first
and second moments of $P$ are identical, which is true for Poisson statistics.
Hence the variance of $F$ becomes:
\begin{equation}
\sigma^2(F)\,=\, \overline{F^2}-\overline{F}^2
           \,=\, \overline{N}\cdot \overline{f^2}
     \,\approx\, \sum_{k=1}^{N}f(D_k)^2.
\end{equation}


\begin{thebibliography}{}
%
\bibitem[1997]{AA}
        Alcock C., Allsman R.A., Alves D. et al. 1997, ApJ 479, 119
\bibitem[1994]{AB}
	Arendt R.G., Berriman G.B., Boggess N. et al. 1994, ApJ 425, L85
\bibitem[1986]{AS}
        Athanassoula E., Sellwood J.A. 1986, MNRAS 221, 213
\bibitem[1983]{BSM}
        Bacon R., Simien F., Monnet G. 1983, A\&A 128, 405
\bibitem[1983]{BSS}
	Bahcall J.N., Schmidt M., Soneira R.M. 1983, ApJ 265, 730
\bibitem[1968]{BN}
        Becklin E.E., Neugebauer G. 1968, ApJ 151, 145
\bibitem[1995]{BSL}
        Beers T.C., Sommer-Larsen J. 1995, ApJS 96, 175
\bibitem[1996]{BGS}
        Binney J., Gerhard O., Spergel D. 1996, preprint astro-ph/9609066,
        submitted to MNRAS
\bibitem[1991]{BG}
        Binney J., Gerhard O.E., Stark A.A., Bally J., Uchida K.I. 1991,
        MNRAS 252, 210
\bibitem[1987]{BT}
        Binney J., Tremaine S. 1987. In: Ostriker J.P. (Ed.)
        Galactic Dynamics. New Jersey, Princeton Univ. Press, p. 120
\bibitem[1996]{BE}
        Bissantz N., Englmaier P., Binney J., Gerhard O. 1996,
        preprint astro-ph/9612026, submitted to MNRAS
\bibitem[1995]{BC}
        Blum R.D., Carr J.S., Sellgren K., Terndrup D.M. 1995, ApJ 449, 623
\bibitem[1978]{BL}
        Burton W.B., Liszt H.S. 1978, ApJ 225, 815
\bibitem[1996]{BU}
        Buta R. 1996. In: Buta R., Crocker D.A., Elmegreen B.G. (eds.)
        Proc. IAU Coll. 157, Barred Galaxies. ASP Conf. Ser. 91, p. 11
\bibitem[1981]{CO}
	Caldwell J.A.R., Ostriker J.P. 1981, ApJ 251, 61
\bibitem[1996]{GK}
        de Grijs R., van der Kruit P.C. 1996, A\&AS 117, 19
\bibitem[1964]{DV}
        de Vaucouleurs G. 1964. In: Kerr F.J., Rodgers A.W. (eds.)
        Proc. IAU-URSI Symp. 20, The Galaxy and the Magellanic Clouds.
        Canberra, Aust. Acad. Sci., p. 88
\bibitem[1996]{DD}
        Durand S., Dejonghe H., Acker A. 1996, A\&A 310, 97
\bibitem[1995]{DA}
        Dwek E., Arendt R.G., Hauser M.G. et al. 1995, ApJ 445, 716
\bibitem[1994]{E}
        Evans N.W. 1994, ApJ 437, L31
\bibitem[1993]{FR}
	Friedli D., Benz W. 1993, A\&A 268, 65
\bibitem[1997]{F}
        Fux R. 1997, PhD thesis, Geneva University
\bibitem[1994]{FM}
	Fux R., Martinet L. 1994, A\&A 287, L21
\bibitem[1996]{FD}
        Fux R., Martinet L., Pfenniger D. 1996. In: Blitz L., Teuben~P. (eds.)
        Proc. IAU Symp. 169, Unsolved problems of the Milky Way.
        Dordrecht, Kluwer, p. 125
\bibitem[1996]{G}
        Gerhard O. 1996. In: Blitz L., Teuben P. (eds.) Proc. IAU Symp. 169,
        Unsolved problems of the Milky Way. Dordrecht, Kluwer, p. 79
\bibitem[1996]{GG}
        Gyuk G. 1996, preprint astro-ph/9607134
\bibitem[1996]{HP}
        Harding P. 1996. In: Morrison H., Sarajedini A. (eds.) Formation of
        the Galactic halo inside and out. ASP Conf. Ser. 92, p. 151
\bibitem[1993]{HL}
        Hernquist L. 1993, ApJS 86, 389
\bibitem[1992]{K}
        Kent S.M. 1992, ApJ 387, 181
\bibitem[1991]{KDF}
	Kent S.M., Dame T.M., Fazio G. 1991, ApJ 378, 131
\bibitem[1994]{KP}
        Kiraga M., Paczynsky B. 1994, ApJ 430, L101
\bibitem[1995]{KB}
        Knapen J.H., Beckman J.E., Heller C.H., Shlosman I., De Jong R.S.
        1995, ApJ 454, 623
\bibitem[1995]{KU1}
        Kuijken K. 1995, ApJ 446, 194
\bibitem[1996]{KU2}
        Kuijken K. 1996. In: Buta R., Crocker D.A., Elmegreen B.G. (eds.)
        Proc. IAU Coll. 157, Barred Galaxies. ASP Conf. Ser. 91, p. 504
\bibitem[1995]{KD}
        Kuijken K., Dubinski J. 1995, MNRAS 277, 1341
\bibitem[1989]{KG}
        Kuijken K., Gilmore G. 1989, MNRAS 239, 605
\bibitem[1989]{LF}
        Lewis J.R., Freeman K.C. 1989, AJ 97, 139
\bibitem[1993]{M}
        Majewski S.R. 1993, ARA\&A 31, 575
\bibitem[1982]{MH}
        Matsumoto T., Hayakawa S., Koizumi H. et al. 1982. In: Riegler G.,
        Blandford R. (eds.) The Galactic Center. New York: AIP, p. 48
\bibitem[1981]{MB}
        Mihalas D., Binney J., 1981. In: Galactic astronomy. San Francisco,
        Freeman Company, p. 525
\bibitem[1996a]{M1}
        Minniti D. 1996a, AJ 112, 590
\bibitem[1996b]{M2}
        Minniti D. 1996b, ApJ 459, 579
\bibitem[1992]{MW}
        Minniti D., White S.D.M., Olszewski E.W., Hill J.M. 1992, ApJ 393, L47
\bibitem[1975]{MN}
        Miyamoto M., Nagai R. 1975, Publ. Astron. Soc. Japan 27, 533
\bibitem[1986]{ML}
        Mulder W.A., Liem B.T. 1986, A\&A 157, 148
\bibitem[1987]{N} 
        Norris J. 1987. In: Gilmore G., Carswell B. (eds) The Galaxy.
        Dordrecht, Reidel, p. 297
\bibitem[1994]{PC}
        Pfenniger D., Combes F., Martinet L. 1994, A\&A 285, 79
\bibitem[1993]{PF}
        Pfenniger D., Friedli D. 1993, A\&A 270, 561
\bibitem[1991]{PS}
        Preston G.W., Shectman S.A., Beers T.C. 1991, ApJ 375, 121
\bibitem[1996]{R}
        Rich R.M. 1996. In: Morrison H., Sarajedini A. (eds.) Formation of
        the Galactic halo inside and out. ASP Conf. Ser. 92, p. 24
\bibitem[1985]{RL}
        Rieke G.H., Lebofsky M.J. 1985, ApJ 288, 618
\bibitem[1995]{RZ}
        Rix H.W., Zaritsky D. 1995, ApJ 447, 82
\bibitem[1977]{RO}
        Rodgers A.W. 1977, ApJ 212, 117
\bibitem[1996]{RR}
        Ruphy S., Robin A.C., Epchtein N. et al. 1996, A\&A 313, 21
\bibitem[1997]{S}
        Sackett P.D. 1997, preprint astro-ph/9608164, accepted by ApJ
\bibitem[1996]{SR}
        Sadler E.M., Rich R.M., Terndrup D.M. 1996, AJ 112, 171
\bibitem[1985]{S1}
        Sellwood J.A. 1985, MNRAS 217, 127
\bibitem[1993]{S2}
        Sellwood J.A. 1993. In: Holt S.S., Verter F. (eds) Back to the Galaxy.
        New York, AIP, p. 133
\bibitem[1988]{SS}
        Sellwood J.A., Sanders R.H. 1988, MNRAS 233, 611
\bibitem[1990]{SW}
        Sharples R., Walker A., Cropper M. 1990, MNRAS 246, 54
\bibitem[1994]{SF}
        Sommer-Larsen J., Flynn C., Christensen P.R. 1994, MNRAS 271, 94
\bibitem[1992]{SJ}
        Spaenhauer A., Jones B.F., Whitford A.E. 1992, AJ 103, 297
\bibitem[1997]{SM}
	Spergel D.N., Malhotra S., Blitz L. 1997, in preparation
\bibitem[1997]{SU}
        Stanek K.Z., Udalski A., Szymanski M. et al. 1997, ApJ 477, 163
\bibitem[1991]{TC}
        te Lintel Hekkert P., Caswell J.L., Habing H.J., Haynes R.F.,
        Norris R.P. 1991, A\&AS 90, 327
\bibitem[1991]{TR}
        Tyson N.D., Rich R.M. 1991, ApJ 367, 547
\bibitem[1994]{U}
        Udalski A., Szymanski M., Stanek K.Z. et al. 1994, Acta Astron. 44,
        165
\bibitem[1995]{V}
        Vall\'ee J.P. 1995, ApJ 454, 119
\bibitem[1995]{VM}
        Verdes-Montenegro L., Bosma A., Athanassoula E. 1995, A\&A 300, 65
\bibitem[1994]{WT}
        Wada K., Taniguchi Y., Habe A., Hasegawa T. 1994, ApJ 437, L123
\bibitem[1981]{W}
        Wamsteker W. 1981, A\&A 97, 329
\bibitem[1994]{WA}
        Weiland J.L., Arendt R.G., Berriman G.B. et al. 1994, ApJ 425, L81
\bibitem[1996]{WS}
        Weiner B.J., Sellwood J.A. 1996. In: Blitz L., Teuben P. (eds.)
        Proc. IAU Symp. 169, Unsolved problems of the Milky Way.
        Dordrecht, Kluwer, p. 145
\bibitem[1977]{WR}
        Wielen R. 1977, A\&A 60, 263
\bibitem[1994]{WO}
        Worthey G. 1994, ApJS 95, 107
\bibitem[1996]{ZX}
        Zhang X. 1996, ApJ 457, 125
\bibitem[1996]{Z}
	Zhao H.S. 1996, MNRAS 283, 149
\bibitem[1996]{ZM}
        Zhao H.S., Mao S. 1996, MNRAS 283, 1197
\bibitem[1996]{ZR}
        Zhao H.S., Rich R.M., Spergel N.D. 1996, MNRAS 282, 175
\bibitem[1985]{ZI}
        Zinn R. 1985, ApJ 293, 424
%
\end{thebibliography}
\end{document}